\documentclass[12pt, onecolumn, letterpaper]{IEEEtran}
\RequirePackage{times}

\RequirePackage[usenames]{pstricks} 
\RequirePackage{graphicx}
\RequirePackage{rotating}
\RequirePackage{indentfirst} 
\RequirePackage{wrapfig}
\RequirePackage{color} 
\RequirePackage{cite} 
\RequirePackage[square,sort&compress,comma,numbers]{natbib}
\RequirePackage{epsfig}
\RequirePackage{subfigure}
\RequirePackage{amsbsy} 
\RequirePackage{amsmath}
\RequirePackage{amssymb}
\RequirePackage{amsthm}
\RequirePackage{latexsym}
\RequirePackage{mathptm} 
\DeclareMathVersion{bold} 
\RequirePackage[final,notref]{showkeys} 
\RequirePackage{bm}
\RequirePackage{multicol}
\RequirePackage{esvect} 
\RequirePackage{ifthen}
\usepackage{setspace}
\usepackage{ulem}
\usepackage{mathptm}
\usepackage{amsmath}
\usepackage{mathptmx}
\usepackage{amsfonts}
\usepackage{dotlessj} 
\usepackage[sectionbib]{chapterbib}
\usepackage{mathtools}

\usepackage{graphicx}
\usepackage[small,bf]{caption}

\newcommand{\ifthen}[2]{\ifthenelse{#1}{#2}{}}

\RequirePackage[hypertex]{hyperref}

\newcommand{\ignore}[1]{}

\newcommand{\COMMENT}[1]{}

\renewcommand{\jmath}{j}

\newcommand{\halmos}{\hskip\textwidth minus\textwidth \rule{6pt}{6pt}}

\newtheorem{lemma}{Lemma}[section]

\newtheorem{theorem}{Theorem}

\newcommand{\eqnlabel}[1]{\label{eq:#1}}
\newcommand{\figlabel}[1]{\label{fig:#1}}
\newcommand{\figref}[1]{Fig.~\ref{fig:#1}}

\newcommand{\seclabel}[1]{\label{sec:#1}}
\newcommand{\secref}[1]{\S\ref{sec:#1}}

\let\er=\eqnref

\newcommand{\eg}{{\it e.g.\/}}
\newcommand{\ie}{{\it i.e.\/}}
\newcommand{\etc}{{\it etc.\/}}

\newcommand{\thmlabel}[1]{\label{thm:#1}}

\newcommand{\numberedtheorem}[3]{
	\ifthen{\not\equal{#1}{}}{
		\let\oof=\thetheorem
		\renewcommand{\thetheorem}{#1}
	}
	\begin{theorem}
	\ifthen{\not\equal{#2}{}}{
		\thmlabel{#2}
	}
	#3
	\end{theorem}
	\ifthen{\not\equal{#1}{}}{
		\let\thetheorem=\oof
		\addtocounter{theorem}{-1}
	}
}
\newcommand{\lemlabel}[1]{\label{lem:#1}}

\newcommand{\numberedlemma}[3]{
	\ifthen{\not\equal{#1}{}}{
		\let\oof=\thelemma
		\renewcommand{\thelemma}{#1}
	}
	\begin{lemma}
	\ifthen{\not\equal{#2}{}}{
		\lemlabel{#2}
	}
	#3
	\end{lemma}
	\ifthen{\not\equal{#1}{}}{
		\let\thelemma=\oof
		\addtocounter{lemma}{-1}
	}
}

\newif\ifjournalsub

\newcommand{\httpref}[1]{}

\newcommand{\be}[1]{\begin{equation}\eqnlabel{#1}}
\newcommand{\ee}{\end{equation}}

\def\longrightharpoonup{\relbar\joinrel\rightharpoonup}
\def\longleftharpoondown{\leftharpoondown\joinrel\relbar}
\def\longrightleftharpoons{\mathop{\vcenter{\hbox{\ooalign{\raise1pt\hbox{$\longrightharpoonup\joinrel$}\crcr\lower1pt\hbox{$\longleftharpoondown\joinrel$}}}}}}

\newcommand{\fREF}{$f_{\text{REF}}$}
\newcommand{\fSYNC}{$f_{\text{SYNC}}$}
\newcommand{\fOSC}{$f_{\text{OSC}}$}

\def\ssp{\def\baselinestretch{0.95645}\large\normalsize}
\def\ssp{\def\baselinestretch{0.7859}\large\normalsize}
\def\ssp{\def\baselinestretch{0.88}\large\normalsize}

\begin{document}
\IEEEoverridecommandlockouts
\title{
    Boolean Computation Using Self-Sustaining Nonlinear Oscillators
}
\author{
    Jaijeet Roychowdhury,
    EECS Department,
    University of California, Berkeley
} 

\maketitle 
\thispagestyle{empty} 

\begin{abstract}
\noindent
    Self-sustaining nonlinear oscillators of practically any type can function as
    latches and registers if Boolean logic states are represented physically as
    the phase of oscillatory signals. Combinational operations on such
    phase-encoded logic signals can be implemented using arithmetic negation and
    addition followed by amplitude limiting. With these,
    general-purpose Boolean computation using a wide variety of natural and
    engineered oscillators becomes potentially possible. Such phase-encoded logic shows
    promise for energy efficient computing. It also has inherent noise immunity
    advantages over traditional level-based logic.
\end{abstract}

\ssp

\section{INTRODUCTION}
\seclabel{introduction}

Self-sustaining oscillators abound in nature and in engineered systems --
examples include mechanical clocks \cite{Huygens:1672}, electronic ring
\cite{FeuerEtAl:1983,Deen:2003,Farzeen:2010} and LC oscillators
\cite{HorowitzHill}, spin-torque oscillators
\cite{KakaEtAl:STNO_Mutual_Locking:2005Natur.437..389K,Mistral:2006,Houss:2007,Devolder:2007,Houss:2008},
lasers \cite{Siegman86,Kobayashi80,IntegratedOpticsTheoryTech2009},
MEMS/NEMS-based oscillators
\cite{NgCICC04,FeWhHaRoSelfSustainingUhfNEMSoscillatorNature2008}, the heart's
neuronal pacemakers \cite{CardiacElectrophysiology2010}, engineered molecular
oscillators such as the repressilator \cite{Elowitz:2000}, \etc. The defining
characteristic of a self-sustaining oscillator is that it generates sustained
``motion'' without requiring any stimulus of a similar nature -- \ie, it
produces an output that changes with time indefinitely, usually in a periodic or
quasi-periodic \cite{Farkas94} fashion, in the absence of any input that changes
with time. If left undisturbed, most practical self-sustaining oscillators
become periodic with time and settle to a single amplitude of oscillation. For
the latter property\footnote{known technically as asymptotic orbital stability
\cite{Farkas94}.}  to hold, the oscillator must be nonlinear, \ie, it must be a
self-sustaining \textit{nonlinear} oscillator (SSNO). SSNOs exhibit interesting
dynamical properties -- for example, synchronization
\cite{StrogatzSync2003,Strogatz00,Strogatz:Stewart:ScientficAmerican:1993} and
pattern formation
\cite{Belousov:1985:reprint,Zaikin:Zhabotinsky:1970:Nature,LaRoASPDAC06nano,BhSrLaRoICCAD2008}
can result when they are coupled together.  Biological phenomena such as the
synchronized flashing of fireflies \cite{Buck76}, circadian rhythms
\cite{SCNsyncModel,Mammalian1} and epilepsy
\cite{KramerKirschSzeriEpilepticPatternFormation2005} result from the
interaction of SSNOs, while coupled systems of SSNOs have been shown to have
image processing capabilities \cite{rick18,LaRoASPDAC06nano} and have been
proposed for associative memories
\cite{NiLaHo2004,RoskaEtAlOscillatoryCNNassociativeMemory2012}. 

In this paper, we first review recent work that establishes that SSNOs can also
serve as substrates for \textit{general-purpose Boolean computation}.  By
exploiting a phenomenon known as sub-harmonic injection locking (SHIL), almost
any SSNO can store logical states stably if logic is encoded in phase.
  This result implies that almost
any oscillator, from any physical domain, can potentially be used for Boolean
computation -- examples include CMOS ring oscillators, spin-torque
nano-oscillators, synthetic biological oscillators, MEMS/NEMS-based oscillators,
nanolasers and even mechanical clocks. Since logic values are encoded in phase,
or time shift, switching between them does not, in principle, involve energy
expenditure. We demonstrate this using a high-Q (energy
efficient) oscillator design that consumes essentially no energy to switch
quickly (in half an oscillation cycle) between phase logic states.
We also outline how phase logic can have generic noise immunity advantages over
level-based encoding of logic.

Phase-encoded logic was first proposed in the 1950s by Eiichi Goto
\cite{Goto:1954,Goto:1955} and John von Neumann
\cite{vonNeumann:1954:NLC,Wigington:1959:ProcIRE}, who showed that if the phase
of a signal (relative to another signal, the reference)
is used to encode Boolean logic states, combinational operations can be
implemented using arithmetic addition and negation. Moreover, they devised a
circuit that served as a phase logic latch -- \ie, it could store a Boolean
logic state encoded in phase.\footnote{This circuit was not, however, a SSNO; it
relied on a sinusoidal (AC) parametric pump (power source) to achieve
bi-stability in phase.}  In the early 1960s, the Japanese constructed phase logic
computers (dubbed Parametrons
\cite{Oshima:1955,Muroga:Parametron:1958,ParametronThocpWebpage,ParametronWikiPedia})
that enjoyed brief success on account of their compactness and reliability
compared to the vacuum-tube based machines that were the mainstay of computing
at the time.
However, phase-based computers were soon overshadowed by level-based ones
employing microscopic semiconductor devices within integrated circuits. The
difficulty of miniaturizing and integrating components in
Goto/von Neumann's phase logic latches contributed to their demise.  Although
subsequently, Goto and colleagues showed that Josephson-junction devices could
be used for phase logic
\cite{Goto:DCfluxParametron:1986,Goto:QuantumFluxParametron:1991}, these require
extremely low temperatures for operation, hence are not practical in most
applications.

With CMOS miniaturization facing fundamental energy and noise barriers today,
there has been an ongoing search for alternative computational
paradigms \cite{ThSoQuestForNextSwitchPROCIEEE2010,ShanbhagEtalDesignAndTest2008}.
In this context, the facts that phase-encoded logic allows essentially
zero-energy bit flips, and is capable of resisting noise better than
level-encoded logic, provide considerable motivation for re-examining
it as a candidate technology for the post-CMOS era.
Also, that any SSNO can potentially serve as a latch removes an important
limitation that prior phase-based logic schemes have faced, \eg, 
many types of nanoscale SSNOs become
candidates for phase latches.\footnote{In this paper, we use CMOS ring
and high-Q LC oscillators for illustration, but other nanoscale SSNOs such as
spin-torque oscillators, NEMS-based oscillators, synthetic biological
oscillators, \etc, can also serve as substrates for SSNO-based phase logic.}

The remainder of the paper is organized as follows. In \secref{SSNOs}, the
concept of encoding logic in phase is outlined and it is shown how SSNOs can
be made to serve as phase logic latches. An example of a state
machine using phase logic is also provided. In \secref{energy}, energy
consumption and speed in SSNO-based phase logic are explored.
The superior noise immunity
properties of phase encoded logic are outlined in \secref{noise}.

\section{Phase logic latches using SSNOs}
\seclabel{SSNOs}

\begin{figure}[htbp]
    \centering{
        \subfigure[Time domain.\figlabel{phaseencodingTD}]{
            \epsfig{file=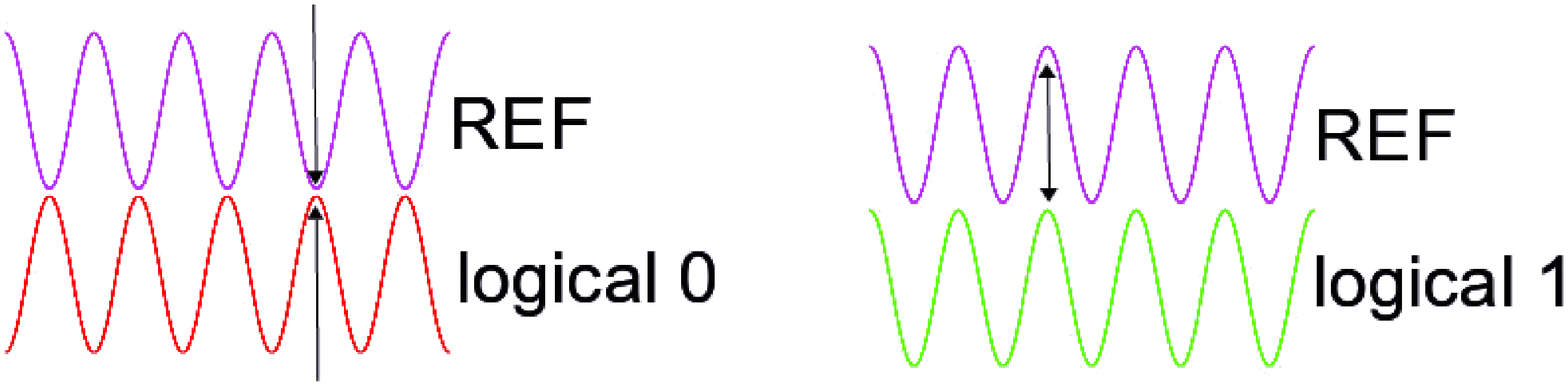,width=0.71\linewidth}
        }
        \subfigure[Phasor.\figlabel{phaseencodingPhasor}]{
            \epsfig{file=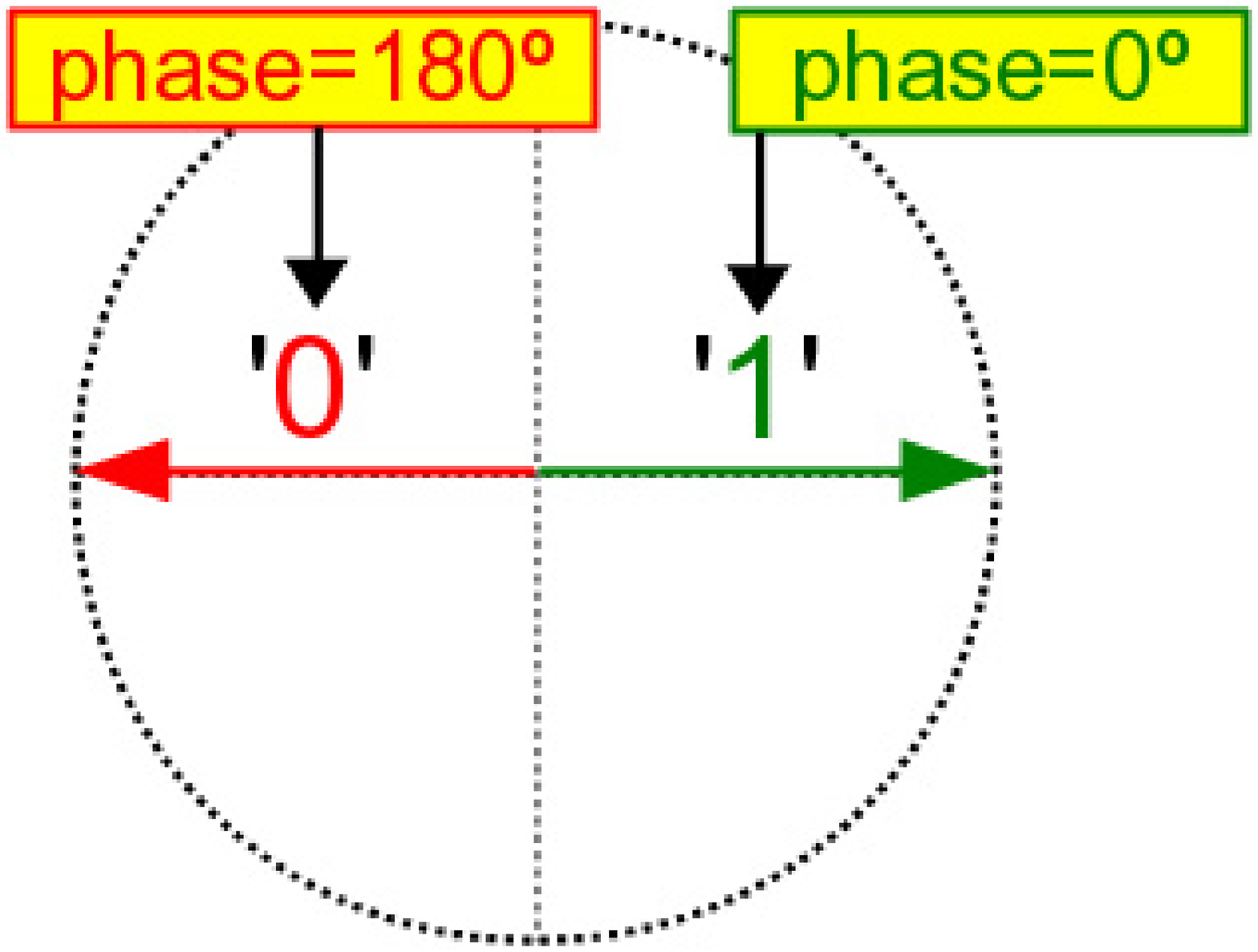,width=0.24\linewidth}
        }
    }
    \caption{\figlabel{phaseencoding}Encoding Boolean logic using the
        relative phase of oscillatory signals.
    }
\end{figure}
\begin{wrapfigure}[14]{r}{0.5\linewidth}
    \vskip-1em
    \centering{
        \epsfig{file=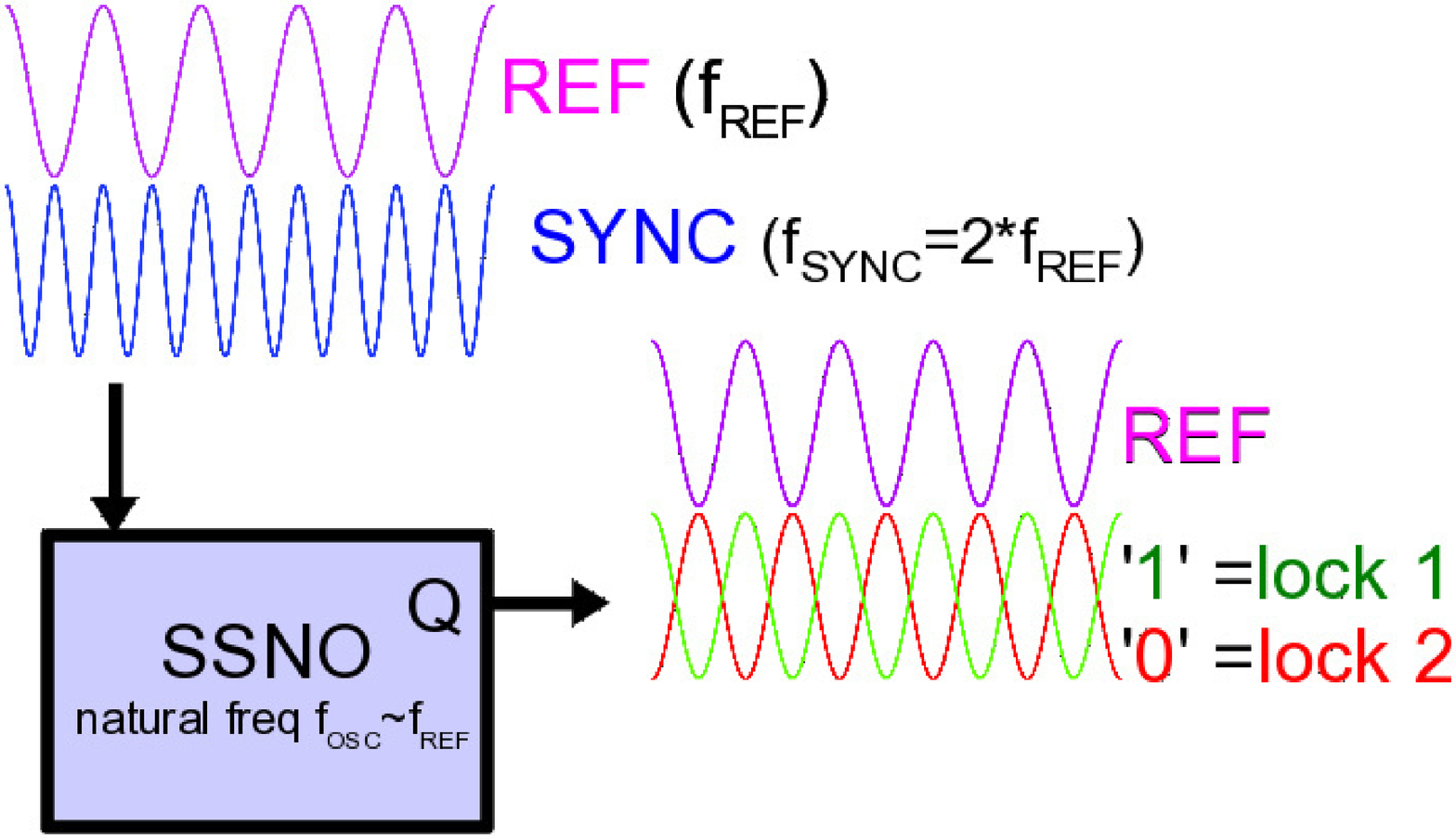,width=\linewidth}
    }
    \vskip-1.5em
    \caption{\figlabel{SSNOphaseLatchScheme}SSNO serving as a bi-stable phase
             latch.}
\end{wrapfigure}
\figref{phaseencoding} above illustrates the use of relative phases to represent
Boolean (binary) logic states.\footnote{Ternary and multi-state logic values can
also be encoded in phase; indeed, SSNOs can serve as multi-state latches
\cite{NeRoDATE2012SHIL}. 
For concreteness, we focus on the binary case
throughout this paper.} A periodic signal, denoted REF in the figure, serves
as a reference with respect to which the phases of other signals are measured.
As shown in \figref{phaseencodingTD}, we choose the opposite phase to represent
logical 0, and the same phase to represent logical 1. Any other choice where the
two logic levels are maximally separated in phase (\ie, by $180^\circ$) would be
equally valid.  Implicit in this scheme is the assumption that all signals
encoding logic using phase are at the same frequency as REF and are
phase locked to it.   The two
phase-encoded Boolean logic states can also be depicted as phasors
\cite{DeCarloLinLinearCktAnalysis}, as shown in \figref{phaseencodingPhasor}.
In the following, we use `1' and `0' to represent the phase-encoded Boolean
states shown in \figref{phaseencoding}.

\subsection{SHIL makes SSNOs phase-bistable}
\seclabel{basicSHIL}
\figref{SSNOphaseLatchScheme} illustrates how an SSNO can be set up as 
a phase logic latch -- \ie, if left undisturbed, it will output either a `1' 
or a `0' (and no other phase) indefinitely in phase synchrony with a provided REF
signal.\footnote{A mathematical proof of this fact for a generic SSNO is
available in \cite{NeRoDATE2012SHIL}.} We assume that
a periodic REF signal with frequency \fREF, as shown
in \figref{phaseencoding} and \figref{SSNOphaseLatchScheme}, is
available. We also require another signal SYNC with frequency exactly
twice that of REF, \ie, \fSYNC$= 2$\fREF.  SYNC is phase-synchronized to
REF, as illustrated in \figref{SSNOphaseLatchScheme}. In practice, SYNC can be
derived from REF by frequency doubling \cite{OzNeAlRFIC2006freqdoubler}, or REF
from SYNC by frequency division \cite{RaLeYaISSCC1994CMOSfreqDivider}.

\begin{figure}[htbp]
    \centering{
        \subfigure[before SHIL\figlabel{SHILillustrationBEFORE}]{
            \epsfig{file=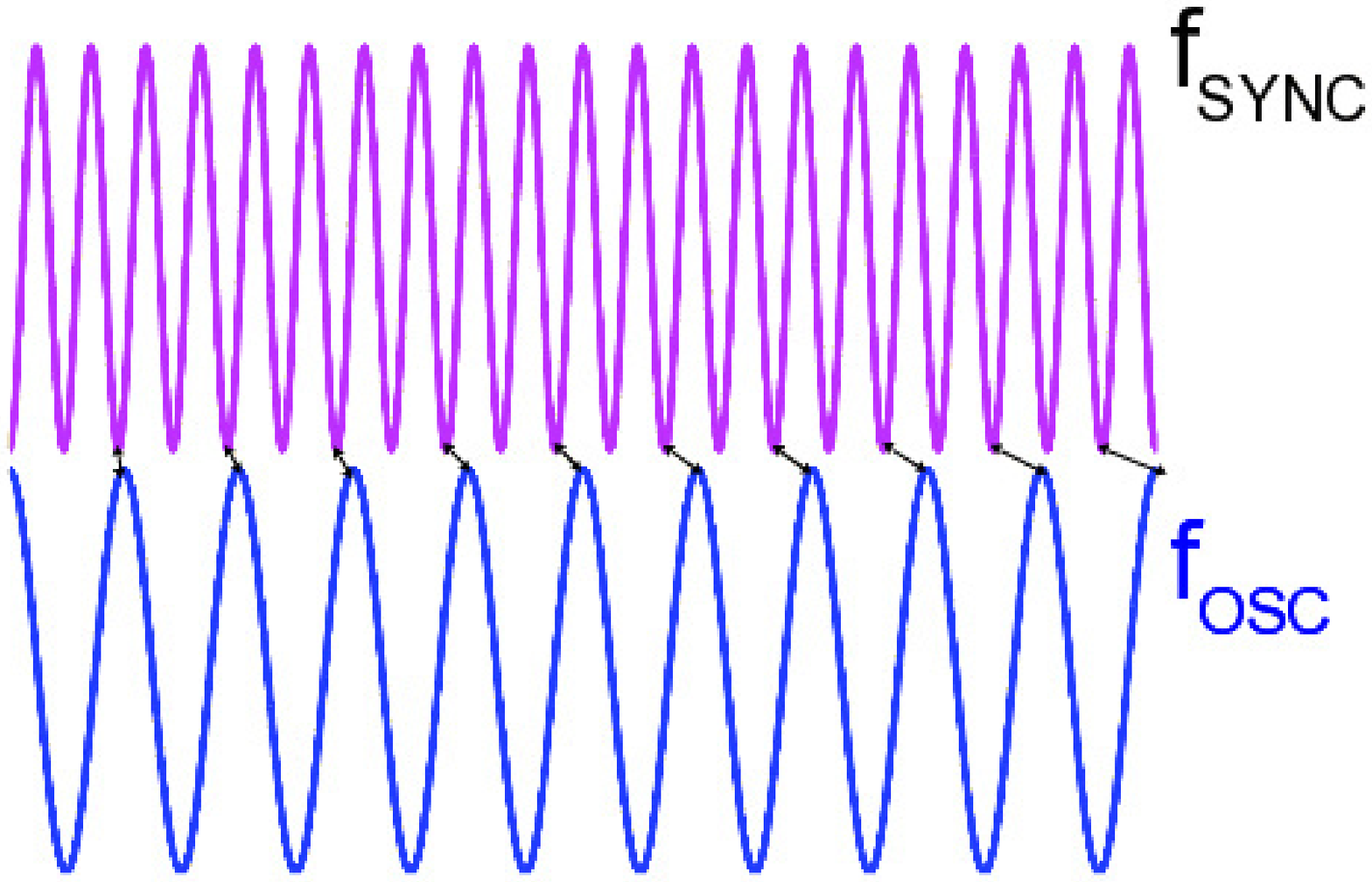,width=0.40\linewidth}
        }\raisebox{5em}{$\Longrightarrow$}
        \subfigure[after SHIL\figlabel{SHILillustrationAFTER}]{
            \raisebox{-1em}{\epsfig{file=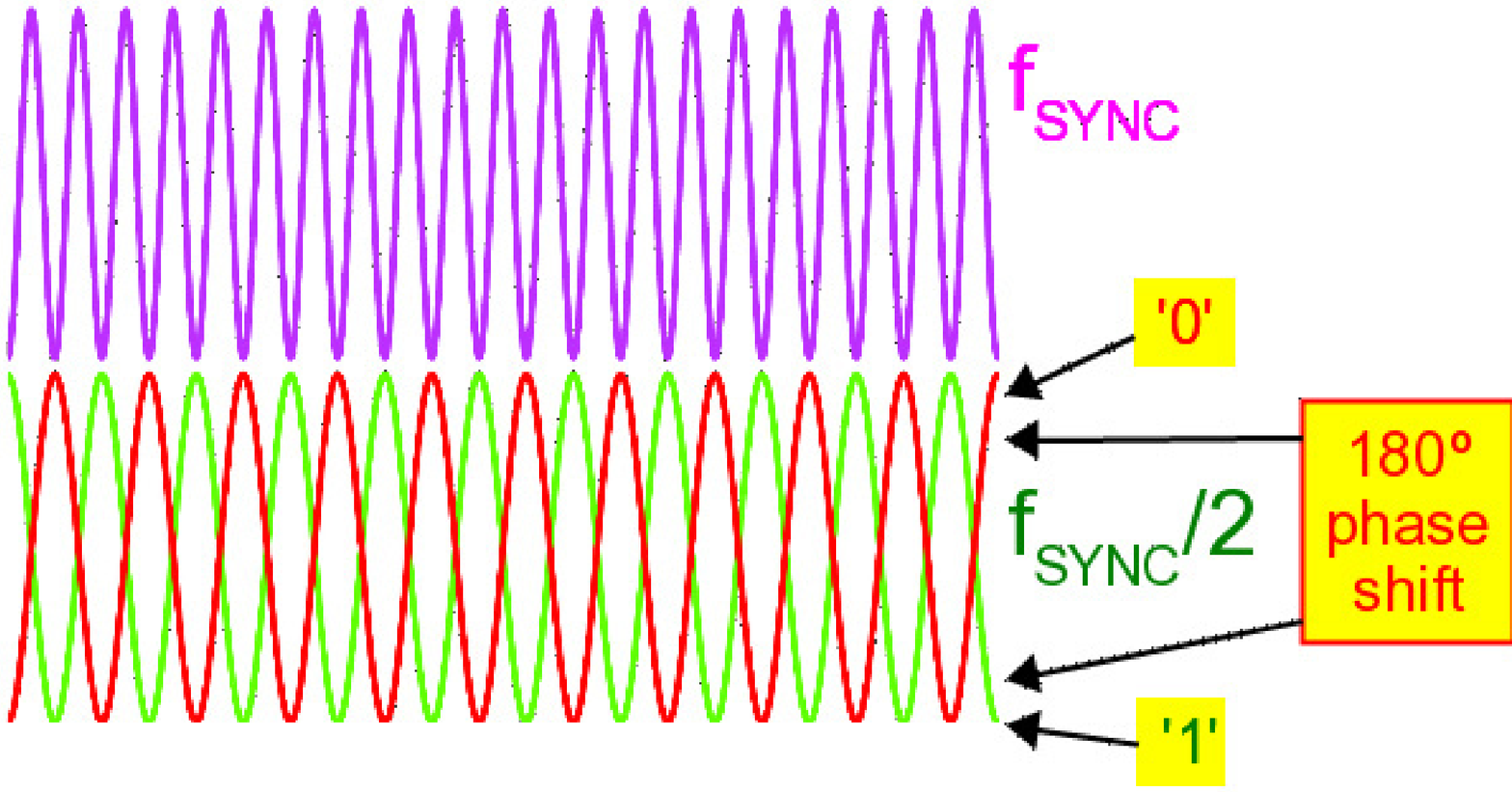,width=0.52\linewidth}}
        }
    }
    \caption{\figlabel{SHILillustration}Sub-harmonic injection locking in an
        SSNO stores phase logic states.
    }
\end{figure}

The SSNO being used as a phase logic latch needs to have a natural frequency
near that of REF, \ie, \fOSC$\simeq$\fREF, or \fOSC$\simeq$\fSYNC/2.
\figref{SHILillustrationBEFORE} illustrates SYNC, juxtaposed against the
oscillator's output at its natural frequency.  Since the oscillator's natural frequency is
only \textit{approximately} half that of SYNC, the two signals are not
necessarily phase synchronized, as depicted by the drift between the two
signals.

The key to devising a phase latch is to \textit{inject the SYNC signal into the
oscillator}, as shown in \figref{SSNOphaseLatchScheme}. With SYNC injection and
under the right conditions \cite{NeRoDATE2012SHIL}, sub-harmonic injection
locking occurs: the oscillator ``forgets'' its natural frequency \fOSC, adopts a
frequency of \textit{exactly} \fSYNC/2, and becomes phase-synchronized with SYNC
in \textit{one of two possible phases that are $180^\circ$ apart}, as depicted
in \figref{SHILillustrationAFTER} by the signals marked `0' and `1'.  In other
words, when SYNC is injected, the oscillator becomes bi-stable in phase at
exactly half the frequency of SYNC and in phase lock with it. That there must be
two stable phase lock states is intuitive because SYNC can ``see'' no difference
between the two lock states (see \figref{SHILillustrationAFTER}); \ie, if the
`0' lock state exists, symmetry dictates that the `1' lock state must also
exist.\footnote{A rigorous proof of SHIL and its bi-stability can be found in
\cite{NeRoDATE2012SHIL}.} Since the oscillator's output is phase locked to SYNC,
it is also phase locked to REF (since SYNC and REF are phase locked by design).
The frequency of the oscillator under SHIL becomes identical to that of REF;
\textit{the key to using the oscillator's two SHIL states for phase logic is
that they can be distinguished using REF.}

\begin{figure}[htbp]
    \centering{
        \subfigure[logic state `0'.\figlabel{MeasurementLogicStateOne}]{
            \epsfig{file=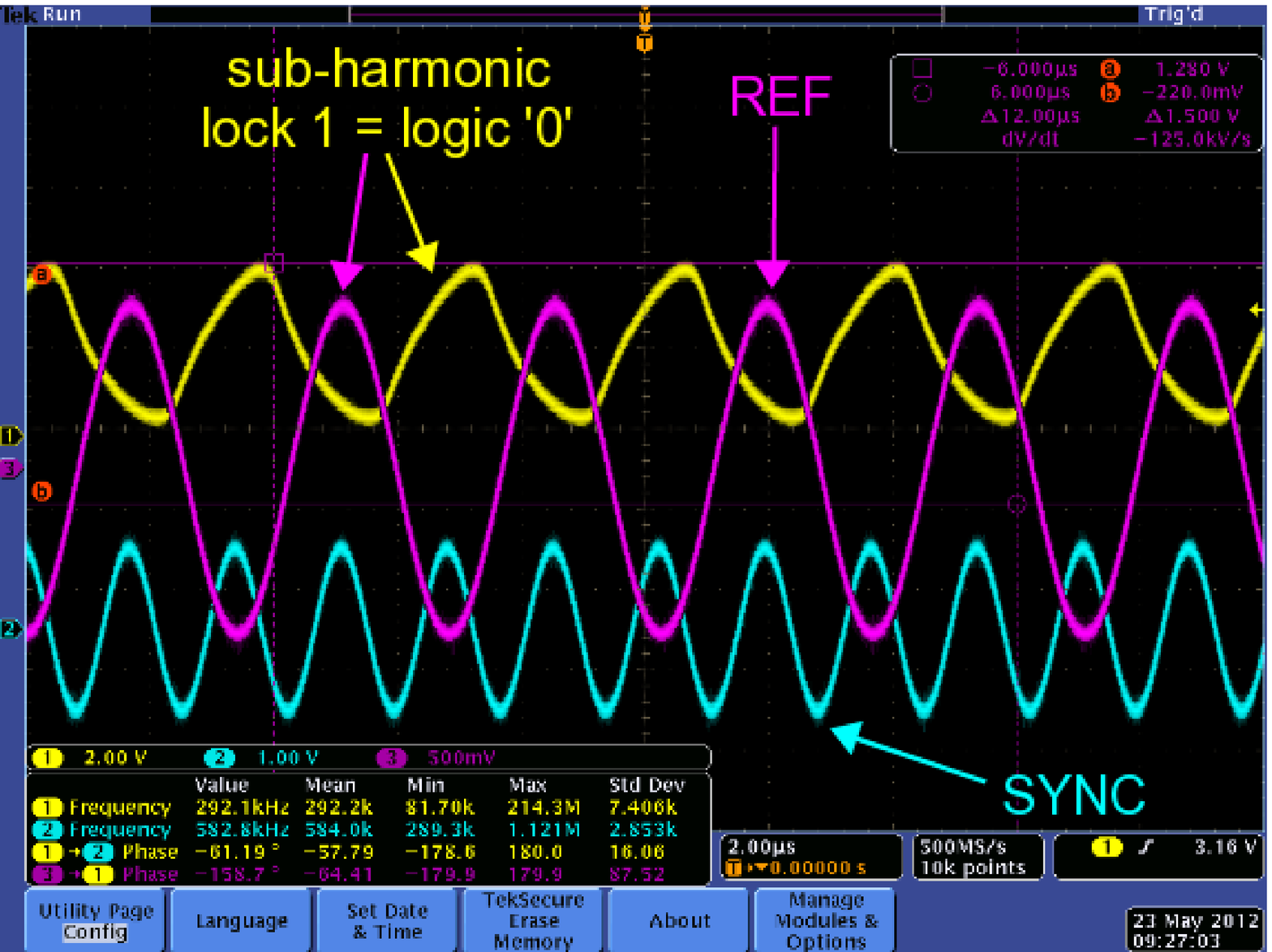,width=0.45\linewidth}
        }
        \subfigure[logic state `1'.\figlabel{MeasurementLogicStateTwo}]{
            \epsfig{file=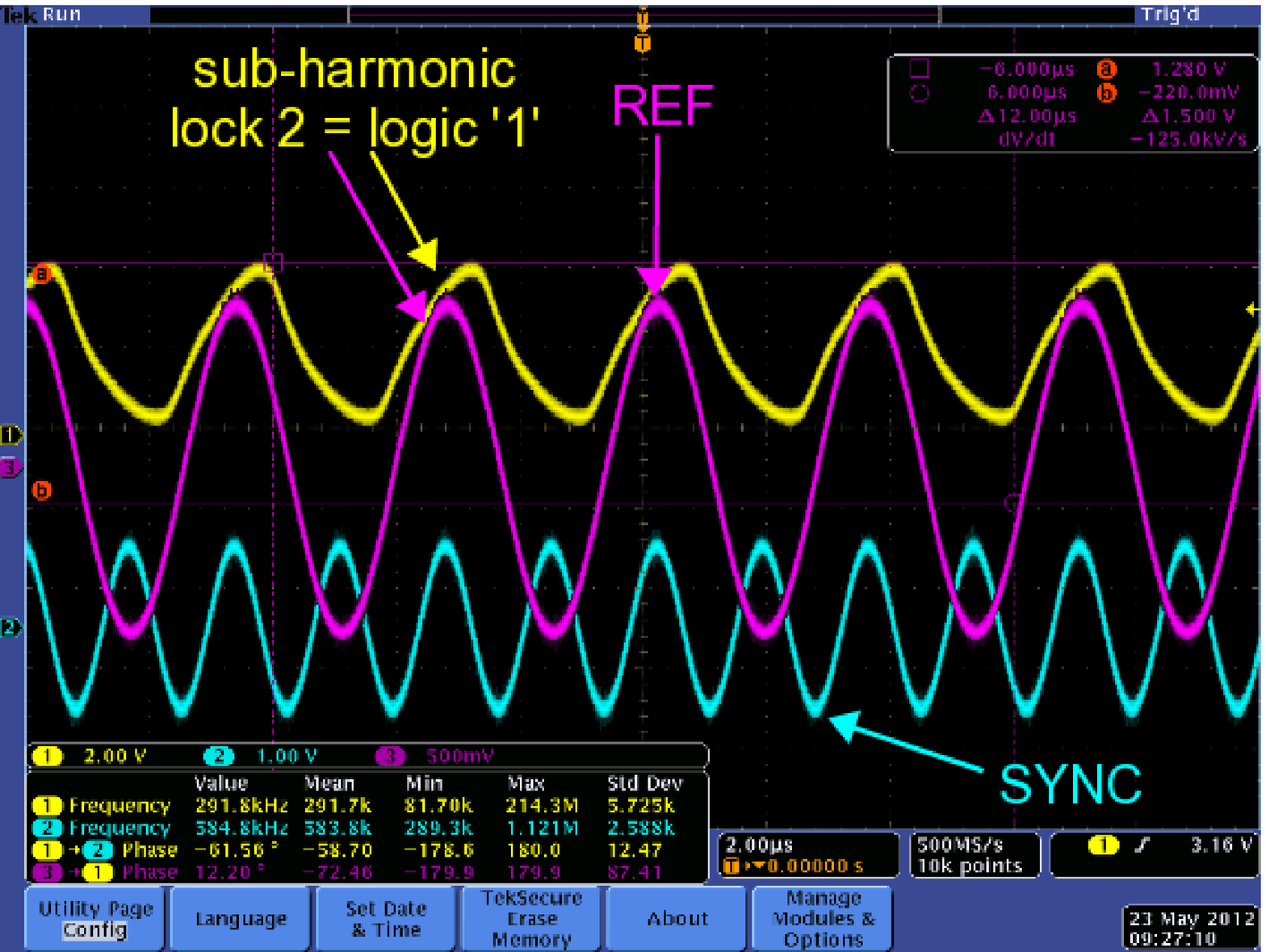,width=0.45\linewidth}
        }
    }
    \caption{\figlabel{Measurements}Oscilloscope traces showing bi-stable SHIL
        in a CMOS ring-oscillator with SYNC injection. 
    }
\end{figure}

Oscilloscope measurements of bi-stable SHIL in a CMOS ring oscillator are shown
in \figref{Measurements}. The SYNC and REF waveforms shown were generated by a
programmable function generator to be in phase lock, with REF at exactly half
the frequency of SYNC. It can be seen that the oscillator's output is at the
same frequency as REF.  In \figref{MeasurementLogicStateOne}, observe that
the peaks of REF are roughly halfway between the peaks of the oscillator's 
output; whereas in \figref{MeasurementLogicStateTwo}, the peaks of REF and
the oscillator's output are almost aligned. These are the two SHIL 
states.\footnote{Which state the oscillator locks to depends on initial
conditions, transients, noise, \etc., 
during circuit startup. \secref{latches} below describes circuits and 
techniques for setting and manipulating the state.}

Using combinational operations, SSNOs featuring bi-stable SHIL can be turned
into D latches \cite{Malvino1995DigitalComputerElectronics}.  We first review
how combinational operations can be implemented using phase logic.

\subsection{Combinational logic in phase}
\seclabel{combinational}
It is well known that certain sets of basic logical operations, when composed,
suffice to implement any combinational logic function. Such sets are called
\textit{logically} or \textit{functionally complete} 
\cite{Wernick1942LogicallyComplete}. For example, the Boolean function sets
\{AND, NOT\}, \{OR, NOT\}, \{NAND\} and \{NOR\} are all logically complete.

When logic is encoded in phase as in \figref{phaseencoding}, it is advantageous
to use the logically complete set \{NOT, MAJ\}
\cite{vonNeumann:1954:NLC,Wigington:1959:ProcIRE}, where NOT is the standard
Boolean inversion operation and MAJ is the 3-input majority operation, returning
whichever Boolean value occurs more than once amongst its three
inputs.\footnote{That \{NOT, MAJ\} is logically complete becomes apparent when
we note that AND(A,B) = MAJ(0, A, B); or that OR(A, B) = MAJ(1, A, B).}  For
example, MAJ(0, 0, 1) returns 0; MAJ(1, 0, 1) returns 1. 

\begin{figure}[htbp]
    \centering{
        \epsfig{file=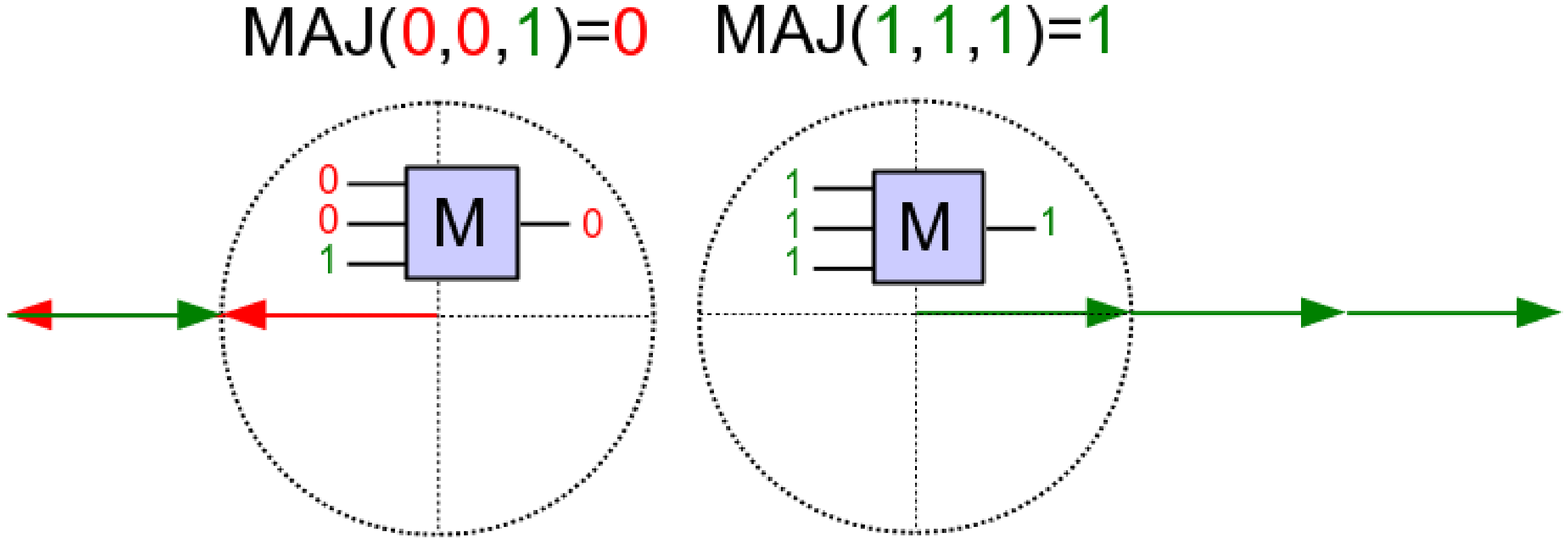,width=0.7\linewidth}
    }
    \caption{\figlabel{MAJbyAddition} Examples illustrating MAJ(A, B, C) in 
              phase logic.}
\end{figure}

The reason \{NOT, MAJ\} is interesting for phase-encoded logic is that both
functions can be implemented using elementary arithmetic operations. NOT
can be implemented simply by arithmetic negation,
as is apparent from \figref{phaseencoding}; it can also be
performed in other implementation-specific ways (\eg, a standard CMOS
inverter topology serves for use with CMOS ring SSNO phase latches; see
\secref{latches} below).  MAJ(A, B, C), where A, B and C are all phase-encoded
logic signals taking values in \{`0', `1'\}, can be implemented by (essentially)
adding A, B and C arithmetically.  This is easy to appreciate graphically using
the phasor representation for phase logic
(\figref{phaseencodingPhasor}), as illustrated using the two examples in
\figref{MAJbyAddition}. Since `0' and `1' are represented by equal and opposite
phasors, adding `0', `0' and `1' leads to the `1' being cancelled by one of the
`0's, leaving `0' -- which is identical to MAJ(`0', `0', `1').  Adding `1', `1',
and `1' results in a phasor with three times the amplitude of `1', but with the
\textit{same phase}; if the amplitude is normalized after addition (\ie, via
amplitude limiting, easily achieved in certain implementations), the result is
`1', which is the same as MAJ(`1', `1', `1').  Arithmetic addition with
amplitude limiting can be confirmed to be identical to MAJ for all other input
combinations.

\subsection{Setting and resetting SSNO SHIL logic states; phase based D-latches}
\seclabel{latches}
\begin{figure}[htbp]
    \centering{
        \subfigure[Generic scheme.\figlabel{SSNOwithLogicInputGeneric}]{
            \epsfig{file=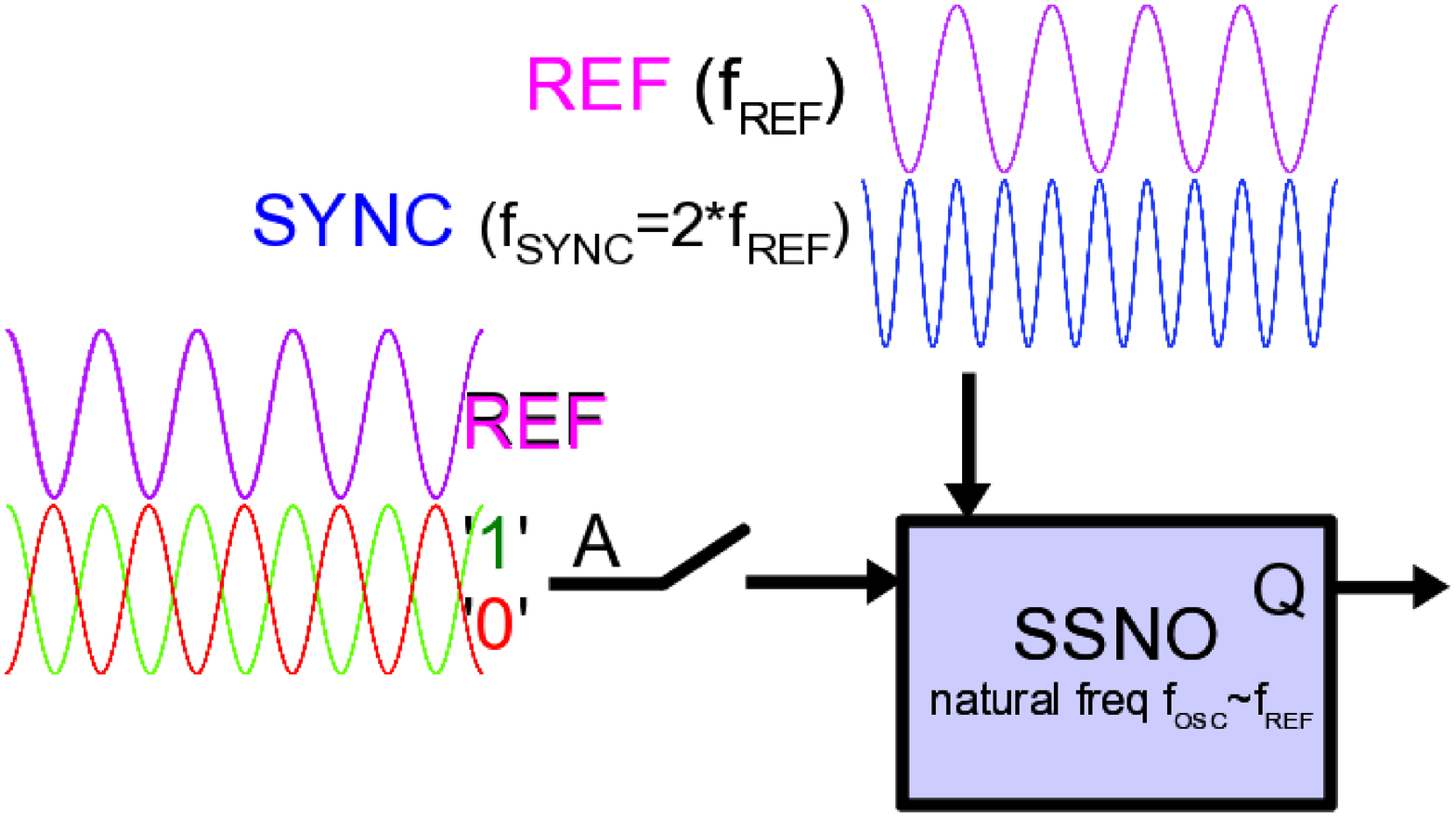,width=0.5\linewidth}
        }
        \subfigure[CMOS ring oscillator example.\figlabel{SSNOwithLogicInputCMOS}]{
            \raisebox{1.5em}{\epsfig{file=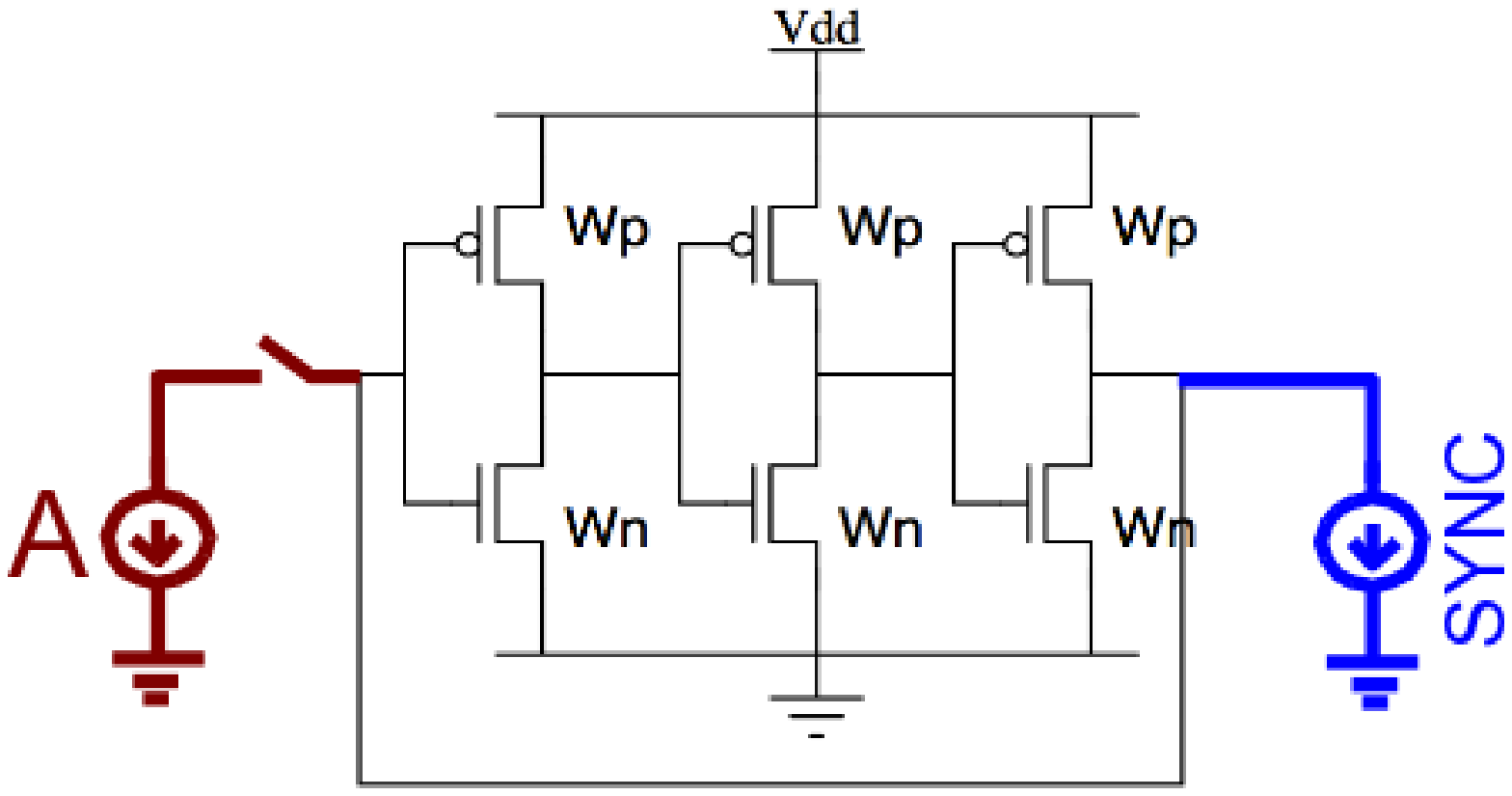,width=0.45\linewidth}}
        }
    }
    \caption{Controlling the lock state of a SSNO under SHIL.\figlabel{SSNOwithLogicInput}}
\end{figure}

To exploit SSNO bi-stability under SHIL (\secref{basicSHIL}) for general
purpose computation, it is necessary to control the SSNO's SHIL state. The basic
mechanism by which this can be achieved is simple, as illustrated in
\figref{SSNOwithLogicInputGeneric}: a phase-encoded logic signal A is injected
into the SSNO momentarily, \eg, by closing the switch briefly. It can be 
shown \cite{WaRoUCNC2014PHLOGON} that under the right circumstances, the SSNO will adopt the
logic state of A and retain it after A is no longer injected.
Injecting the phase-encoded logic signal A (which is at the frequency of REF)
removes SHIL bi-stability under SYNC injection and sets the oscillator's phase
close to that of A \cite[Figure 4]{WaRoUCNC2014PHLOGON};\footnote{This happens
because ``simple'' (\ie, fundamental harmonic) injection locking
\cite{Adler46,LaRoMTT2004,BhRoASPDAC2009}, in which the oscillator becomes phase
locked to A with exactly one stable state, overrides SHIL.} when A is removed,
bi-stability is restored and the oscillator adjusts its phase smoothly to the
nearest SHIL stable lock state, \ie, that of A.

\begin{figure}[htbp]
    \centering{
        \epsfig{file=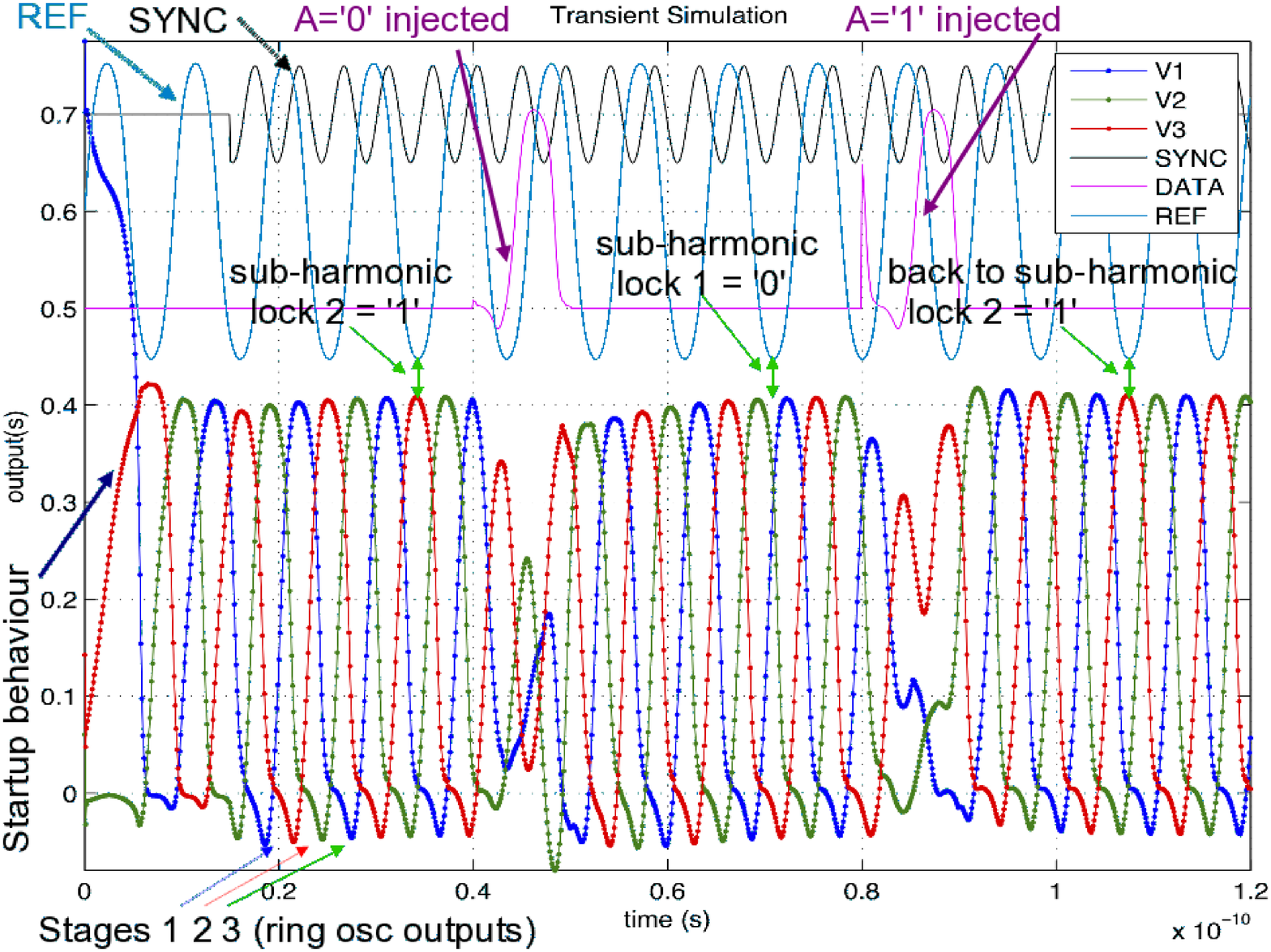,width=0.95\linewidth}
    }
    \caption{Transient simulation of the circuit in
             \figref{SSNOwithLogicInputCMOS}.\figlabel{SHILtransient}}
\end{figure}

\figref{SSNOwithLogicInputCMOS} shows a CMOS ring SSNO with SYNC and A
injections -- the two current injections are at the same node in this case,
though they can be incorporated in a variety of alternative ways. The dynamics
of setting and
resetting the SSNO's SHIL state can be seen in the transient simulation plots
in \figref{SHILtransient}. The first cycle of the ring oscillator's output shows
startup transients in the absence of SYNC injection. SYNC injection starts
at $t\sim 17.5$ps (see the waveform labelled SYNC). The oscillator
responds within about 2 cycles by changing its frequency to \fSYNC/2 and
settling to an arbitrary phase logic
state -- in this case `1', indicated by the oscillator's stage 2 (red) output's
peaks being almost aligned with REF's troughs. At
$t\sim 40$ps, about 1 cycle of A=`0' is injected momentarily (see the
label \texttt{A='0' injected}); the
oscillator's waveforms change significantly in response.
By about $t\sim 70$ps, the oscillator settles to the other
bi-stable SHIL state, \ie `0', as seen by the fact that the trough of REF
is no longer aligned with the oscillator's stage 2 (red) output's peaks, but is 
instead roughly halfway between the peaks. The SHIL state is then switched back to
`1' by momentarily injecting A=`1' at $t\sim 80$ps; the oscillator responds by
switching back to phase logic state `1' by $t\sim 110$ps, with the stage 2
output's peak aligned again with the trough of REF.

\begin{figure}[htbp]
    \centering{
        \epsfig{file=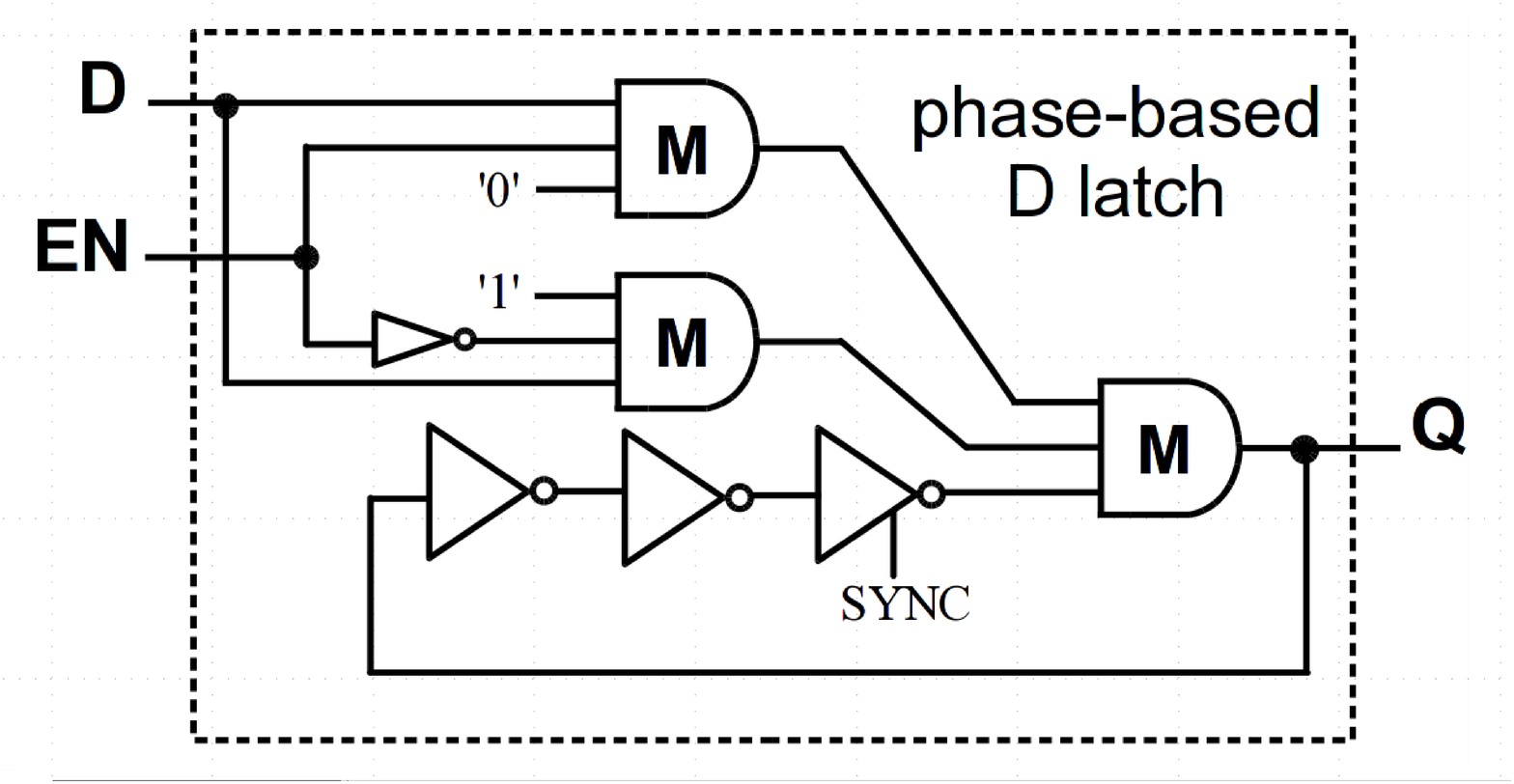,width=0.7\linewidth}
    }
    \caption{\figlabel{Dlatch}Phase based D latch with enable using a CMOS ring
                              oscillator \cite{WaRoUCNC2014PHLOGON}. The gates 
                              marked M are 3-input majority gates.}
\end{figure}

The basic ring oscillator phase latch topology of
\figref{SSNOwithLogicInputCMOS} can be easily adapted \cite{WaRoUCNC2014PHLOGON}
into a gated D latch (D latch with Enable)
\cite{Malvino1995DigitalComputerElectronics} with the help of the combinational
primitives \{NOT, MAJ\}, as shown in \figref{Dlatch}. The chain of three
inverters represents the CMOS SSNO of \figref{SSNOwithLogicInputCMOS} with the
SYNC injection included, but without the input A; direct feedback from the last
inverter to the first is broken and a MAJ gate introduced, as shown.  All logic
I/Os (D, EN, and Q) are phase encoded. The inverter driven by the EN (Enable)
input, representing logical inversion (using phase encoding), can be implemented
simply as a standard CMOS inverter.  When EN=`1', D is fed to two inputs of the majority
gate in the ring oscillator loop, resulting in the ring oscillator's feedback
loop being broken and Q being set to D. When EN=`0', D is ignored and
complementary logic values are fed to two inputs of the majority gate in the
ring oscillator loop, which sets Q to the output of the third inverter in the
ring oscillator (which is the third input to the majority gate), thereby
completing the ring oscillator's feedback loop, restoring bi-stability and
retaining the previously set state.

\subsection{State machines using SSNO-based phase logic}
\seclabel{statemachine}

\begin{figure}[htbp]
    \centering{
        \epsfig{file=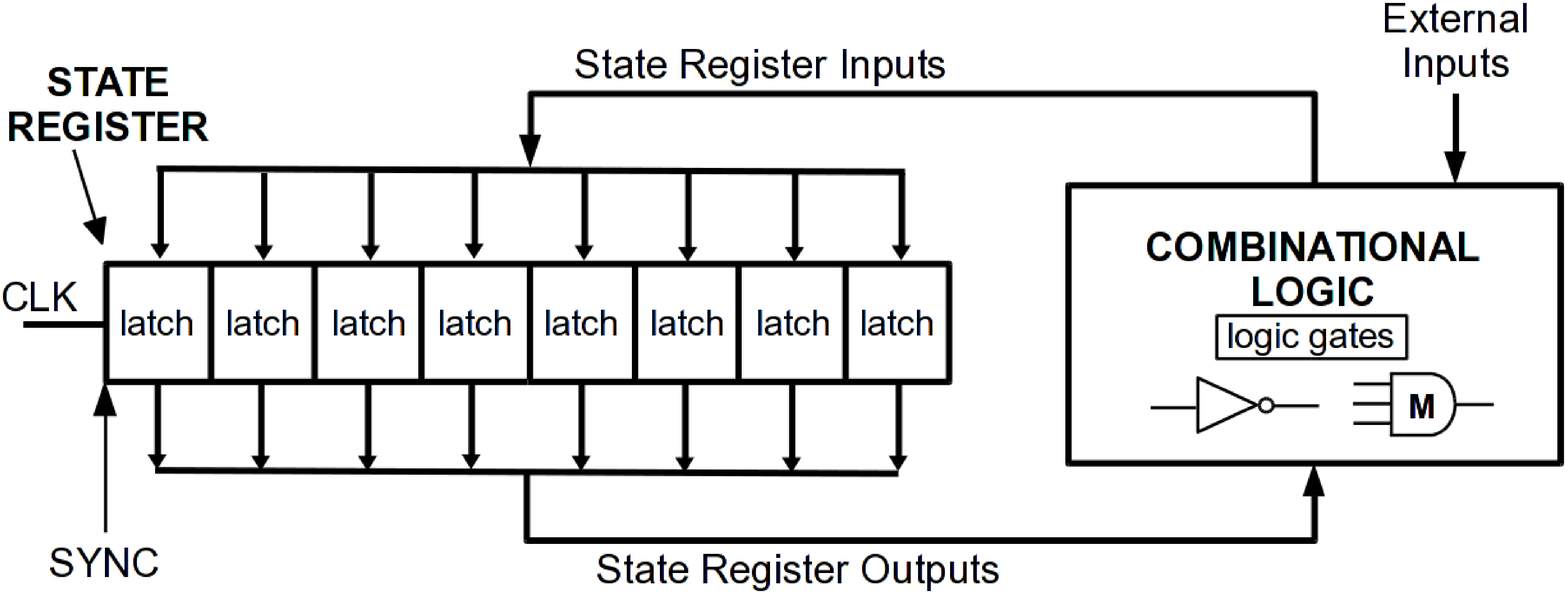,width=0.7\linewidth}
    }
    \caption{\figlabel{GenericStateMachine}Structure of a state machine using
        SSNO-based phase latches and \{NOT, MAJ\} based combinational logic.}
\end{figure}

\begin{wrapfigure}[30]{r}{0.5\linewidth}
    \vskip-1.0em
    \centering{
        \subfigure[State Transition Graph.\figlabel{FullAdderSMSTG}]{
            \epsfig{file=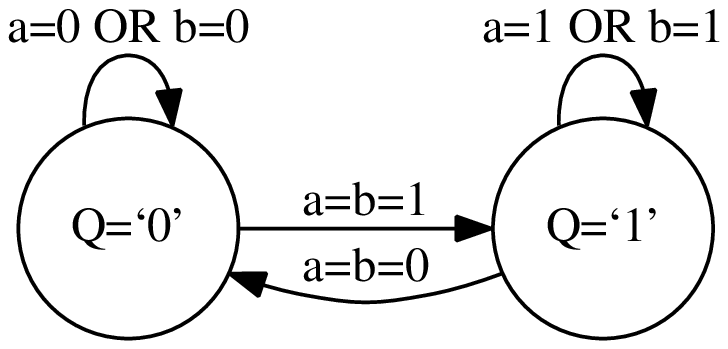,width=0.7\linewidth}
        }\\
        \subfigure[Implementation with phase D latches and \{NOT,
                   MAJ\} gates.\figlabel{FullAdderSMimpl}]{
            \raisebox{0em}{\epsfig{file=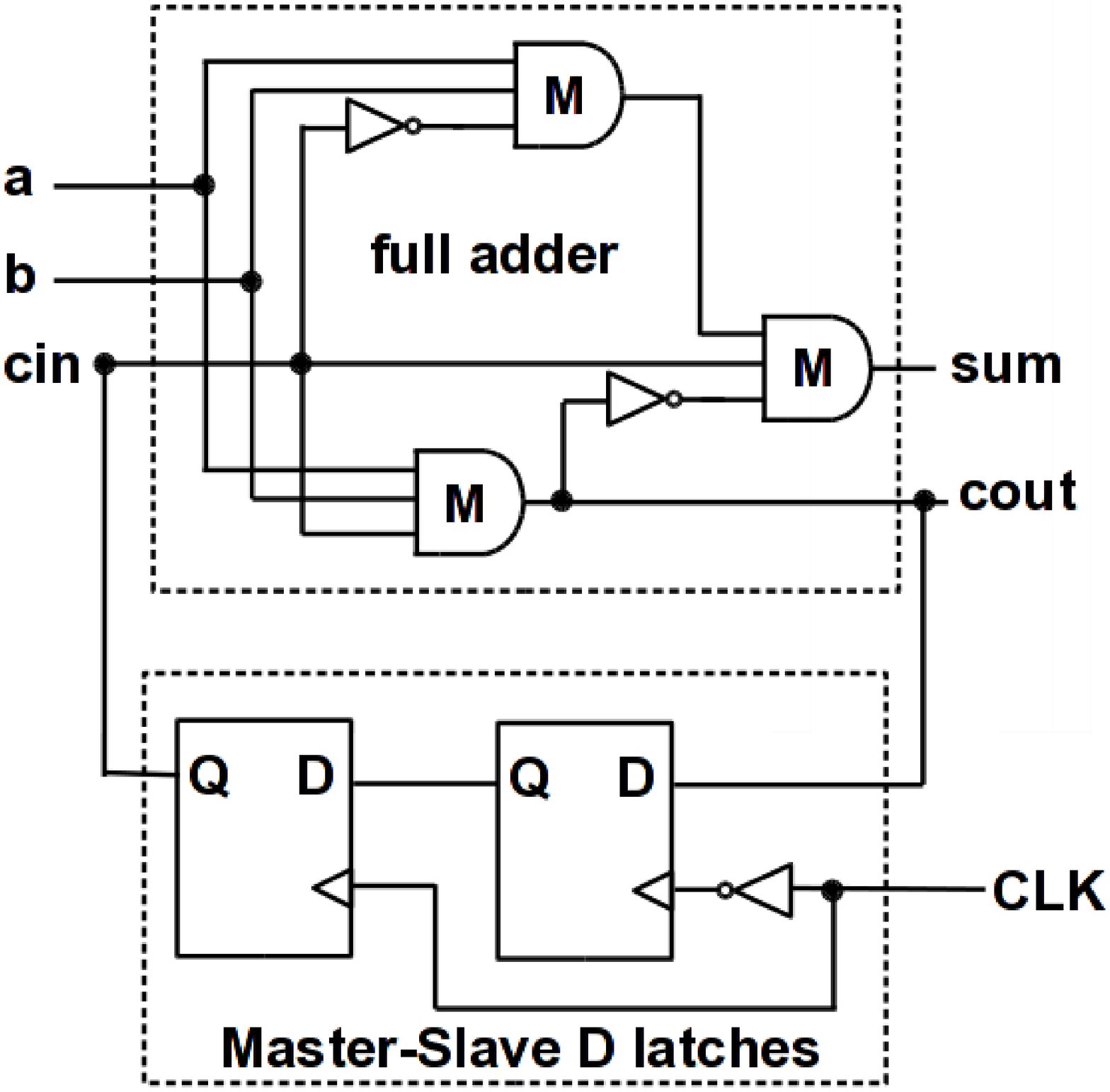,width=0.8\linewidth}}
        }
    }
    \caption{\figlabel{FullAdderSM}SSNO-based 1-bit state machine example
             \cite{WaRoUCNC2014PHLOGON}.}
\end{wrapfigure}
With D latches for storage and combinational logic using \{NOT, MAJ\}, we have
the basic components for a von Neumann computer \cite{vonNeumann:1945:FDR} in
SSNO-based phase-encoded logic.  One of the most important units of a computer
is the finite state machine (FSM), used for, \eg, the control unit
\cite{vonNeumann:1945:FDR,Malvino1995DigitalComputerElectronics} of a stored
program computer.  The general structure of a state machine, adapted to the
phase logic context, is shown in \figref{GenericStateMachine}. All signals are
phase encoded, including the CLK signal which alternates between phases `0' and
`1', holding each for a few cycles of REF. It is also easy to devise D latches
where CLK (ENable) is level-based, while the logic signals remain encoded in
phase; however, a level-based clock signal will not benefit from the increased
noise immunity of phase-based encoding (see \secref{noise}, below).

\figref{FullAdderSMimpl} shows an example of a simple Mealy FSM that utilizes a
full adder for the combinational logic and a single bit for the state, all in
phase logic \cite{WaRoUCNC2014PHLOGON}. The latch is constructed using two of
the D latches shown in \figref{Dlatch}, arranged in a master-slave
\cite{Malvino1995DigitalComputerElectronics} configuration to prevent races.
The two inputs to the state machine, \textbf{a} and \textbf{b}, are inputs to
the full adder. The carry-out (\textbf{cout}) bit of the full adder is the input
to the latch; the output of the latch feeds back as the carry-in (\textbf{cin})
input of the full adder. This arrangement implies that the `0'$\rightarrow$`1'
state transition can only occur if \textbf{a}=\textbf{b}=`1', and the
`1'$\rightarrow$`0' transition if \textbf{a}=\textbf{b}=`0'.  The complete state
transition diagram of the FSM is shown in \figref{FullAdderSMSTG}.

\section{Energy efficiency and switching speed of phase-encoded logic}
\seclabel{energy}

Having outlined the fundamental design and operational principles of SSNO-based
phase logic, we now explore two fundamental questions: how much energy does it
take to flip a bit in phase-encoded logic, and how quickly can a bit be flipped?

\subsection{Energy dissipation and amplitude/phase change rates in high-Q
oscillators}

For reference, the minimum energy expended by level based logic (for which a
single inverter serves as an exemplar) in flipping a bit from 0 to 1 and back
again to 0 is $C V_{\text{DD}}^2$,\footnote{where $C$ is the capacitive load at
each inverter and $V_{\text{DD}}$ is the supply voltage.} averaging
$\frac{1}{2}C V_{\text{DD}}^2$ per bit flip. Just to maintain oscillation,
a minimum energy of $3 C V_{\text{DD}}^2$ is dissipated per cycle by the
3-stage ring oscillator of the previous section, hence it is not a
compelling candidate for energy efficient computation.  Although
dissipation can be lowered using small supply voltages,\footnote{Ring oscillators
operating at 100mV using standard CMOS technologies have been reported
\cite{Deen:2003}.} it is typically at the cost of decreased oscillation
frequency and logic switching speed. 

However, high-Q LC oscillators (\eg,
\cite{VittozEtAlHighPerfCrystalOscsJSSC1988,NgHoJSSC1999}) are inherently
energy efficient, dissipating only about $\frac{1}{Q}$ of the energy stored in
the LC tank\footnote{$\frac{1}{2}C V_{\text{osc}}^2$, where $V_{\text{osc}}$ is
the peak amplitude of oscillation.} per cycle, where $Q$ is the quality factor
of the oscillator.  LC oscillators are also capable of very high frequency
oscillation -- \eg, a $300$ GHz LC oscillator has been reported
\cite{Razavi300GHzFundJSSC2011}. These characteristics make high-Q LC oscillators
interesting candidates for exploring how energy efficient, and how fast,
phase-encoded logic can be.

Using a proof-of-concept circuit, we show that it is possible
to make high-Q LC oscillators suitable for phase logic by subjecting them to
SHIL, and to \textit{flip their phase logic states in just half a cycle
with no energy consumption} (beyond the small amount of energy needed per cycle
to maintain oscillation). Phase logic can therefore be about $Q$ times more
energy efficient than level based flipping, without compromising switching
speed.\footnote{indeed, possibly at far higher speeds than level based logic is
currently capable of, depending on the oscillator's frequency.} With Q
factors of $10^2$-$10^6$ readily achievable,
great energy savings over level based logic can potentially result.

\begin{wrapfigure}[15]{r}{0.26\linewidth}
    \vskip-1em
    \centering{
        \epsfig{file=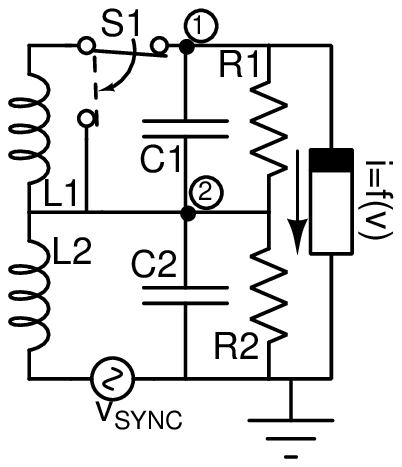,width=\linewidth}
    }
    \vskip-1.5em
    \caption{\figlabel{TwoTankOscLatch}High-Q LC oscillator based
                        phase logic latch circuit.}
\end{wrapfigure}
Being able to flip a bit (\ie, disturb one normal oscillation pattern
and settle to another) within a single cycle of a high-Q oscillator may appear
counter-intuitive, since amplitude changes in high-Q oscillators are very slow on
account of their necessarily involving energy dissipation or
accumulation in the LC tank.  This energy can be removed or supplied only in
small installments per cycle in high-Q oscillators, translating to slow
amplitude transients with time constants of the order of $Q$ cycles of
oscillation.

However, flipping a phase-encoded
logic bit involves only time shifting or delaying oscillatory waveforms.
There appears to be no fundamental physical principle dictating a minimum energy
needed to achieve a time shift -- therefore, in principle, phase-encoded bit
flipping would seem achievable with no energy consumption at all. With no
need to supply or remove energy, the speed at which time shifts can be made
would seem limited only by the time constants of the oscillatory dynamics of the
LC tank. Since the LC tank changes phase by 360$^\circ$ as a matter of course
during each cycle of oscillation, it should be possible to shift phase by
180$^\circ$ (\ie, to the other stable SHIL phase lock state) in half a
cycle.\footnote{That the slowness limitation of amplitude changes in high-Q LC
oscillators does not apply to their phase/time shifting characteristics
appears not to be widely appreciated.} Our experiments below confirm this
reasoning and provide proof of the concept that zero energy bit flips
can be achieved in half a cycle of a
high-Q LC oscillator.

\subsection{High-Q LC oscillator based phase logic latch}

\figref{TwoTankOscLatch} depicts the schematic of a high-Q LC oscillator
that serves as a phase logic latch. The circuit is based on the
standard parallel-RLC tank and nonlinear resistor topology
\cite{LaWaRoCICC2005}. The oscillator's main tank is the upper one, consisting of
$L_1$, $C_1$ and $R_1$; it is tuned to a natural frequency \fOSC\/, set close to
\fREF\/ = \fSYNC/2. The single-pole double-throw (SPDT) switch S1, normally kept
closed in the position shown, is used for phase logic bit flipping, as
described below.

To facilitate sub-harmonic injection locking, a second tank consisting of $L_2$,
$C_2$ and $R_2$ is tuned to \fSYNC\/ and placed in series with the main tank.
The negative resistance nonlinearity is connected across both tanks, as shown.
The SYNC signal is injected as a voltage source in series with $L_2$. The
second tank magnifies the effect of SYNC on the nonlinear resistor
by a (typically large) factor of $\frac{R_2}{2\pi f_{\text{SYNC}} L_2}$,
thereby sensitizing the oscillator to SHIL from the SYNC signal.

The nonlinearity needs a negative differential resistance region to 
power the circuit and enable self oscillation.  It has the current-voltage
characteristic
\be{fOFv}
    i = f(v) \triangleq k_1 \tanh(k_2 v) + g_{\text{SHIL}}(v),
\ee 
where 
\be{gSHILv}
    g_{\text{SHIL}}(v) \triangleq \begin{cases}
                                   k_3^2 (v+A)^2 & \text{if } v < -A, \\
                                   0 & \text{if } -A \le v \le A, \\
                                   k_3^2 (v-A)^2 & \text{if } v > A.
                                  \end{cases}
\ee
The $\tanh(\cdot)$ term in \er{fOFv} provides the
negative differential resistance needed for oscillation \cite{LaWaRoCICC2005}.
The $g_{\text{SHIL}}(v)$ term facilitates second sub-harmonic
injection locking by introducing asymmetry in $f(v)$ for input amplitudes larger
than $A$. Such asymmetry enables second-harmonic components of
the voltage input to $f(v)$ to affect the phase of the fundamental component of
its current output. Describing function based feedback analysis
\cite{LaWaRoCICC2005} shows that this feature is important for
susceptibility to injection locking. 

\begin{figure}[htbp]
    \centering{
        \subfigure[Main tank voltage (w SYNC overlaid).
                            \figlabel{oscUnderSHILlockVoltages}]{
            \epsfig{file=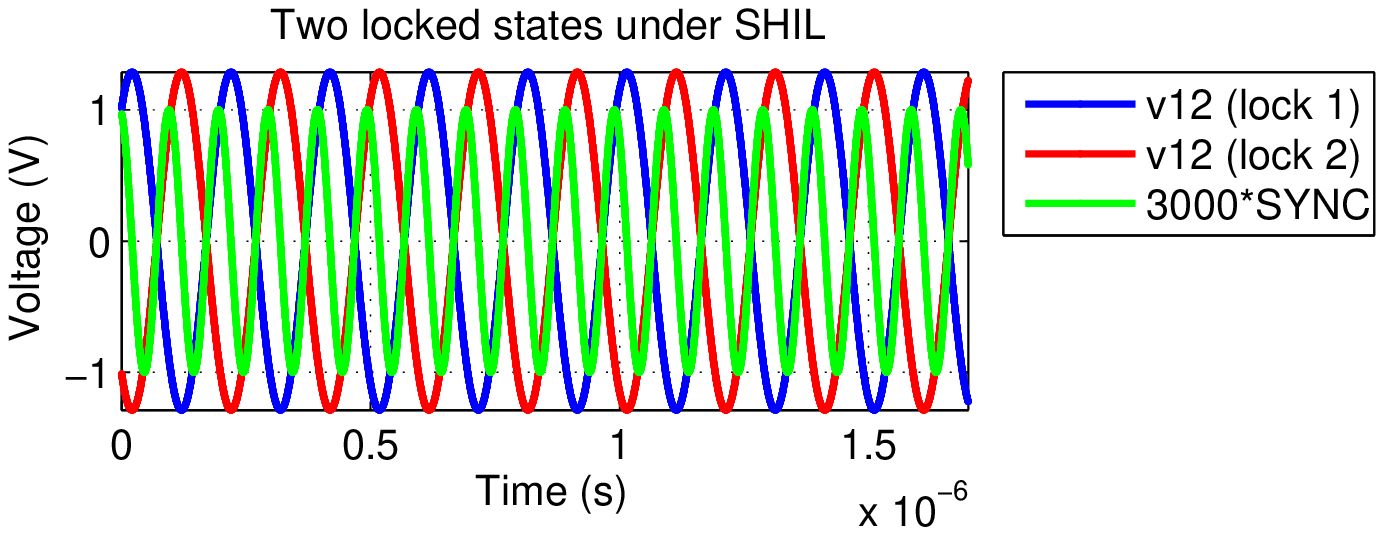,width=0.47\linewidth}
        }
        \subfigure[Instantaneous power.\figlabel{oscUnderSHILlockPower}]{
            \epsfig{file=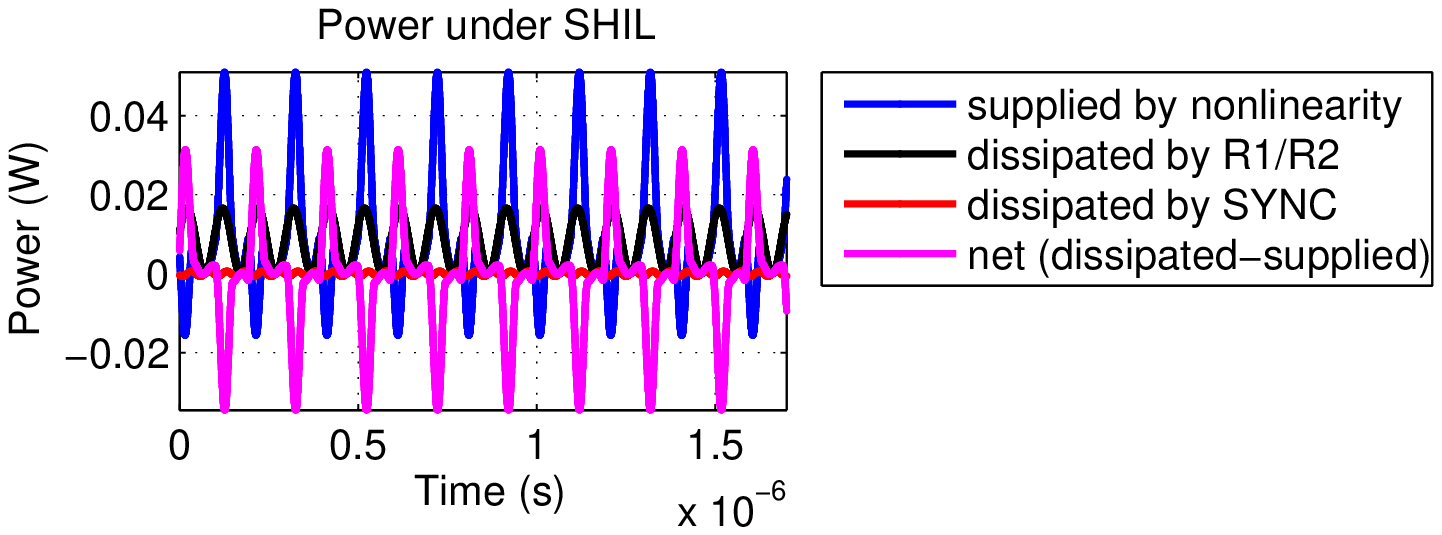,width=0.47\linewidth}
        }
        \subfigure[Cumulative energy.
                            \figlabel{oscUnderSHILlockEnergies}]{
            \epsfig{file=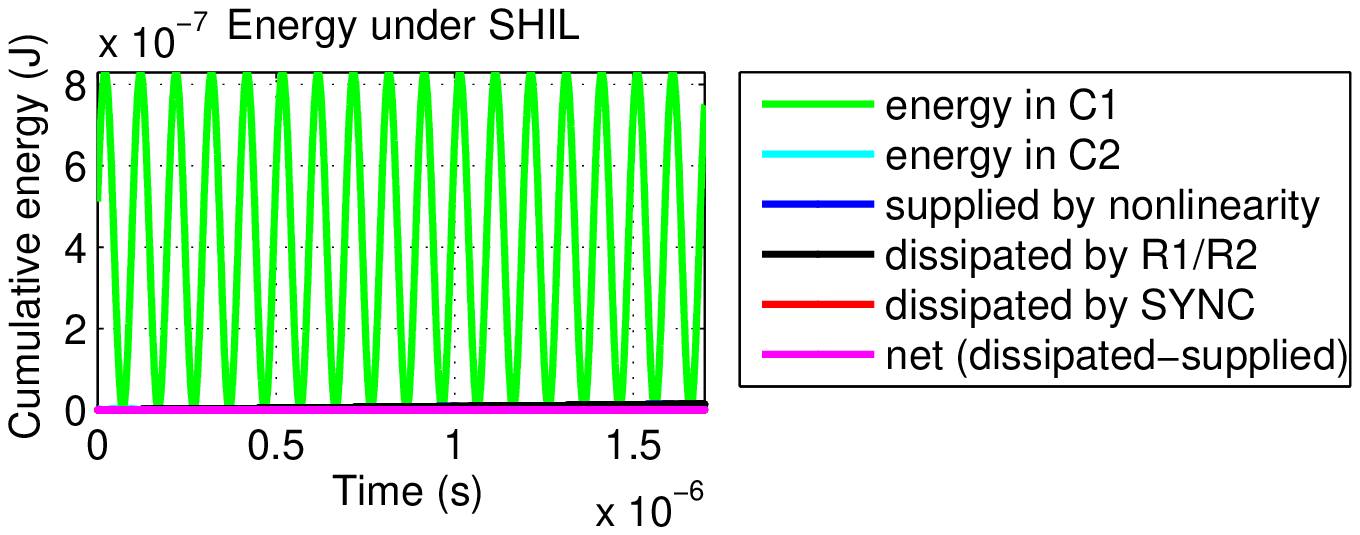,width=0.47\linewidth}
        }
        \subfigure[Cumulative energy (detail).\figlabel{oscUnderSHILlockEnergiesBig}]{
            \epsfig{file=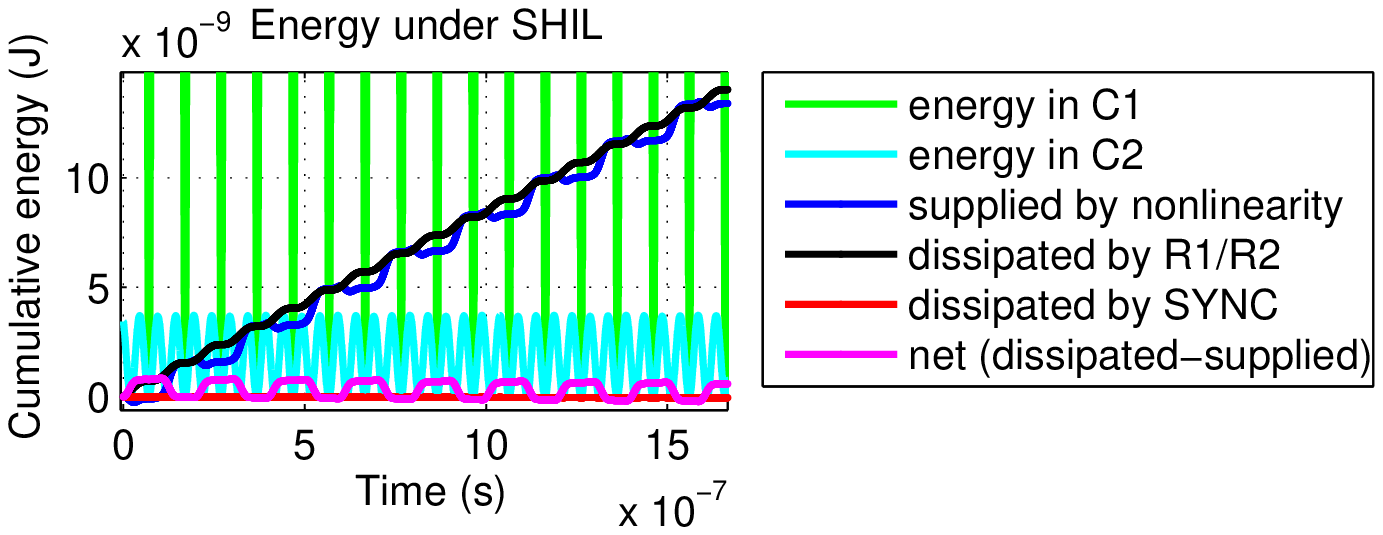,width=0.47\linewidth}
        }
    }
    \caption{\figlabel{oscUnderSHILlock}Voltages, power and energy consumption
                                        of LC oscillator under SHIL.
    }
\end{figure}

For natural oscillation to occur (in the absence of any injection
at $v_{\text{SYNC}}$), it is necessary for 
\be{oscConditions}
    \frac{1}{R_2} > - k_1 k_2 > \frac{1}{R_1},
\ee
\ie, the maximum negative differential resistance of $f(v)$ needs to overcome
the loss due to $R_1$, but not the loss due to $R_2$ --- the latter condition
prevents fundamental-mode natural oscillation at \fSYNC.
The parameter $A$ in \er{gSHILv} is set at or around the amplitude of natural
oscillation (\ie, in the absence of SYNC injection).

The simulations below use the following values of circuit parameters:
\be{parameters}
    \begin{aligned}
    L_1=1\text{nH}, \: C_1=1\mu\text{F},\:  R_1=100\Omega,\:  L_2=\frac{L_1}{2}, 
    \: C_2=\frac{C_2}{2}, \\ R_2 = 90\Omega, \: k_1=\frac{1}{30},\:  
    k_2 = \frac{0.0102}{k_1}, \: k_3 = 40 k_1 k_2, \: A = 0.9.
    \end{aligned}
\ee
The switch $S_1$ was modelled with on resistance $0\Omega$ and off resistance
$10$k$\Omega$. With these parameters, \fOSC$\, \simeq \frac{1}{2\pi \sqrt{L_1
C_1}} \sim 5.03292$ MHz. \fREF\/ was taken to be $5.0328$ MHz, with \fSYNC$\, =
2\,$\fREF. The SYNC injection was 
\be{vSYNCoft}
    v_{\text{SYNC}}(t) = 10^{-3}k_1 \cos(2 \pi f_{\text{SYNC}} \, t).
\ee

\figref{oscUnderSHILlockVoltages} shows the voltage of the main tank of the
oscillator under SHIL.\footnote{All results are from simulation using MAPP
\cite{WaAaWuYaRoCICC2015MAPP, MAPPwebsite}. Harmonic Balance (HB) 
\cite{UsCh84,kundertwhite}
was used to find the two locked steady states; transient simulations were
initialized with the HB solutions.} The two locks, representing logic levels `0'
and `1', can be seen to be exactly 180$^\circ$ out of phase, as predicted by
theory \cite{NeRoDATE2012SHIL,WaRoUCNC2014PHLOGON}. 

The instantaneous power of the various components of the circuit are
shown in \figref{oscUnderSHILlockPower}.  Power is supplied to
the circuit by the nonlinearity $i=f(v)$, and dissipated primarily by the
tank losses $R_1$ and $R_2$.  The SYNC injection signal can also dissipate or
supply power, while the resistances of the switch $S_1$ dissipate power, but 
these amounts are negligible. \figref{oscUnderSHILlockEnergies} and
\figref{oscUnderSHILlockEnergiesBig} depict cumulative energies (\ie, integrated
power) supplied/dissipated by the components; also overlaid are the instantaneous
energies of the tank capacitors $C_1$ and $C_2$, the peak values of which
represent the total energy stored in each tank. The peak value for $C_1$
indicates that the energy of the main tank is about $0.829\mu$J.

\begin{figure}[htbp]
    \centering{
        \subfigure[Main tank voltage overlaid on `0'/`1' waveforms.
                            \figlabel{oscBitFlipsVoltages}]{
            \epsfig{file=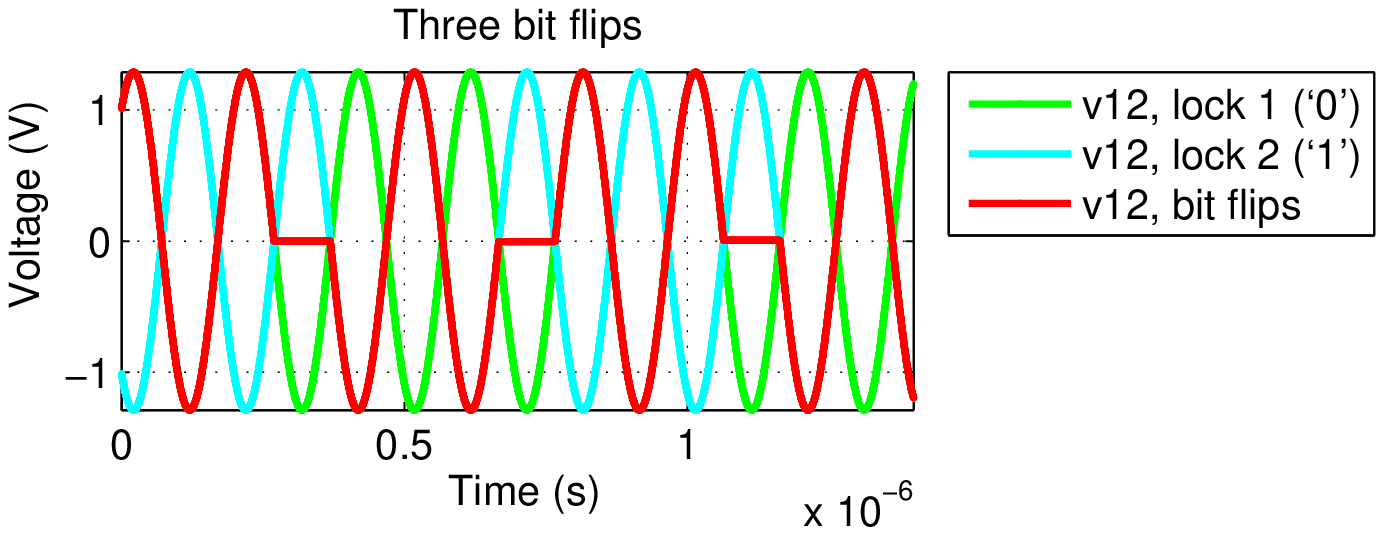,width=0.47\linewidth}
        }
        \subfigure[Instantaneous power.\figlabel{oscBitFlipsPower}]{
            \epsfig{file=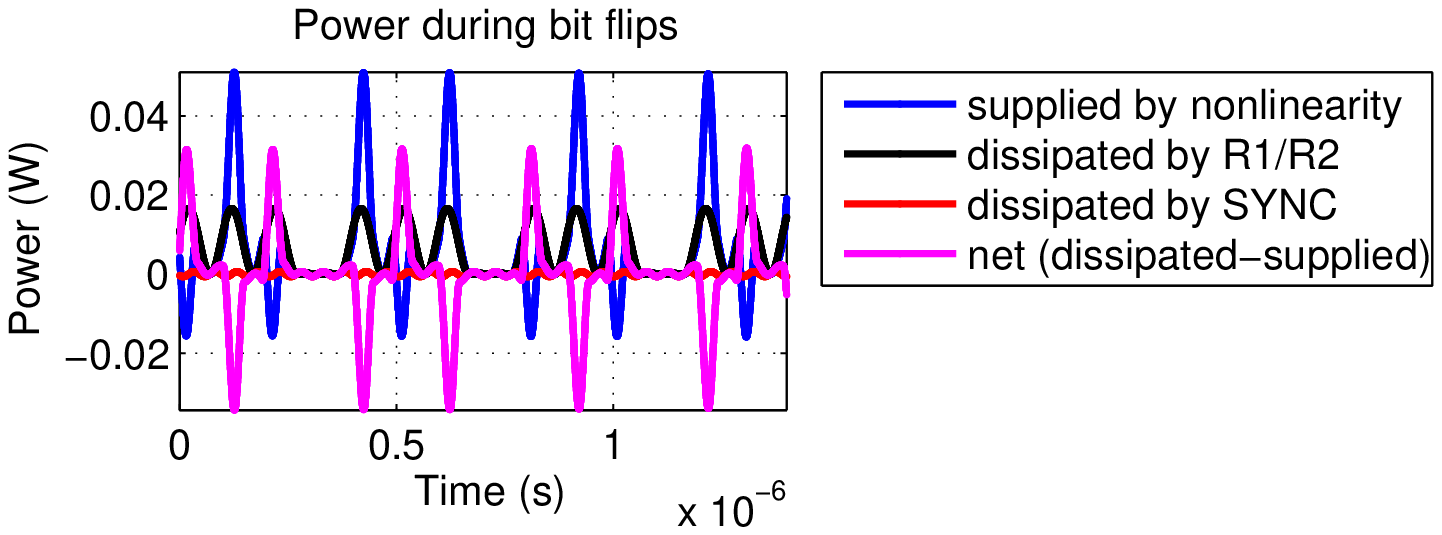,width=0.47\linewidth}
        }
        \subfigure[Cumulative energy.
                            \figlabel{oscBitFlipsEnergies}]{
            \epsfig{file=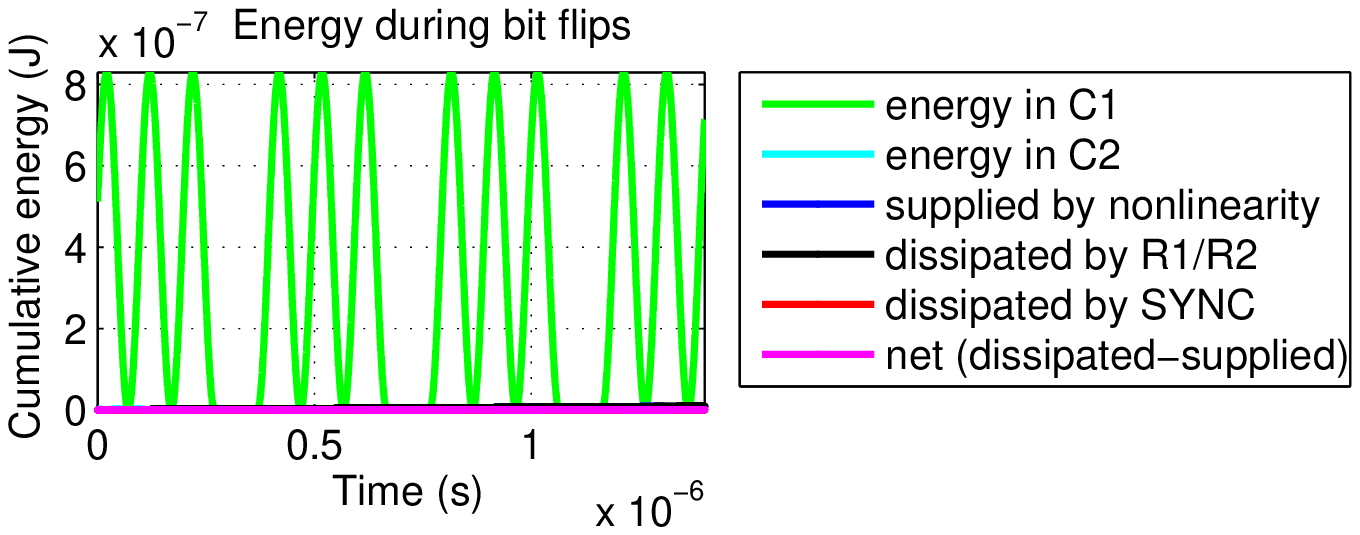,width=0.47\linewidth}
        }
        \subfigure[Cumulative energy (detail).\figlabel{oscBitFlipsEnergiesBig}]{
            \epsfig{file=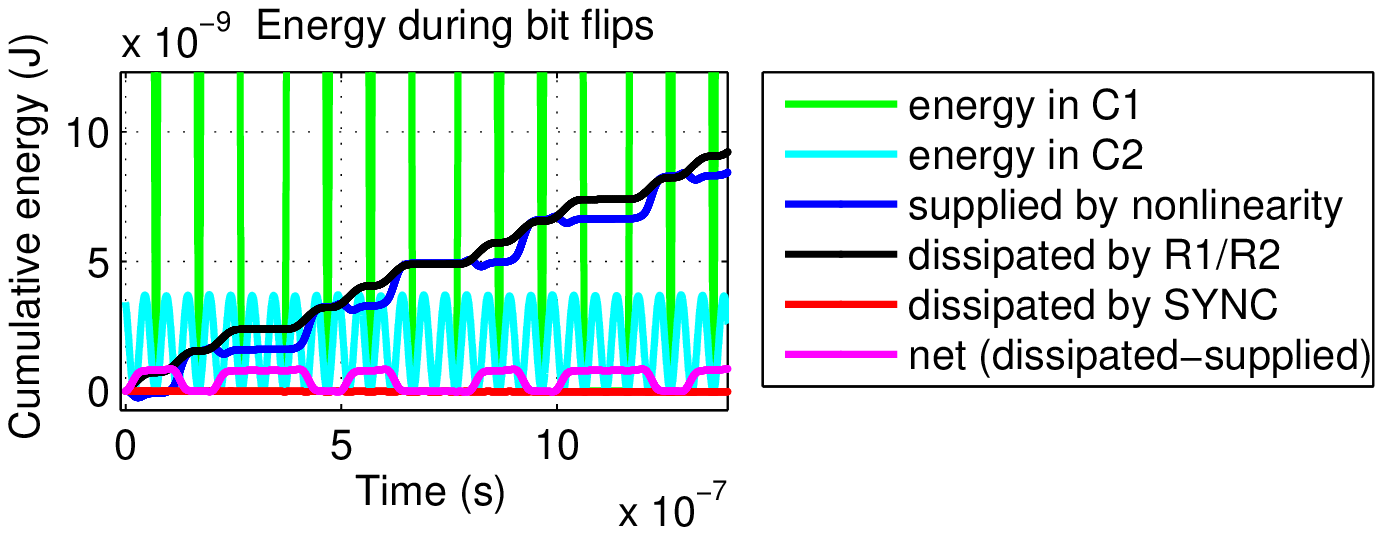,width=0.47\linewidth}
        }
    }
    \caption{\figlabel{oscBitFlips}Voltages, power and energy consumption
                                        of LC oscillator undergoing bit flips.
    }
\end{figure}

As expected in periodic lock, the energy supplied by the nonlinearity
during each cycle exactly compensates the energy dissipated (primarily by the
tank losses) -- the net energy trace in
\figref{oscUnderSHILlockEnergiesBig} periodically crosses zero, implying
that no energy is being gained or lost by the tanks. The
energy supplied to (and dissipated by) the oscillator over each cycle is seen to
be about $1.685$nJ, implying an effective Q factor\footnote{Because of the
nonlinear resistor, the Q of a self-sustaining oscillator is typically lower --
by about 6$\times$ in this case -- than the ideal Q factor of the linear tank
alone \cite{FoundationsOfAandD}.} of about $492$ in sub-harmonically
injection locked operation.

\subsection{Speed and energy during bit flips}

The SPDT switch $S_1$ in \figref{TwoTankOscLatch} can be used to transition the
oscillator between the two SHIL states shown in
\figref{oscUnderSHILlockVoltages}. If $S_1$ is
flipped to short the inductor $L_1$ when the voltage across it is zero, the main
tank's dynamics are frozen in time until $S_1$ is flipped back.  Flipping the
switch for half an oscillation cycle delays the tank just enough to move the
oscillator from one lock state to the other.

The simulation results in \figref{oscBitFlips} illustrate this technique of
achieving phase logic bit flips.\footnote{It is also possible to use other
techniques, such as voltage or current injections, to flip the oscillator's
state.} $S_1$ is flipped for half a cycle three times (starting around
0.27$\mu$s, 0.67$\mu$s, and 1.06$\mu$s), leading to three bit flips.
\figref{oscBitFlipsVoltages} shows the voltage across the main tank overlaid on
the two lock states of \figref{oscUnderSHILlockVoltages}, illustrating how well
the bit flips from each state to the other. Power waveforms are shown in
\figref{oscBitFlipsPower}, while cumulative energies are shown in
\figref{oscBitFlipsEnergies} and \figref{oscBitFlipsEnergiesBig}.  Energy
consumption during bit flipping is small, since the oscillator is
essentially stopped when the bit is being flipped. Similar to
\figref{oscUnderSHILlockEnergiesBig}, the net energy graph in
\figref{oscBitFlipsEnergiesBig} crosses zero after bit flipping, indicating that
no energy is being gained or lost by the tanks. This shows that the energy
benefits due to the high Q of the oscillator are reaped even as bits are flipped at
high speed (in half an oscillation cycle).

\section{Noise immunity of phase-encoded logic}
\seclabel{noise}
\begin{figure}[htbp]
    \centering{
        \subfigure[Small noise amplitude.\figlabel{levelbasednoiseSmall}]{
            \epsfig{file=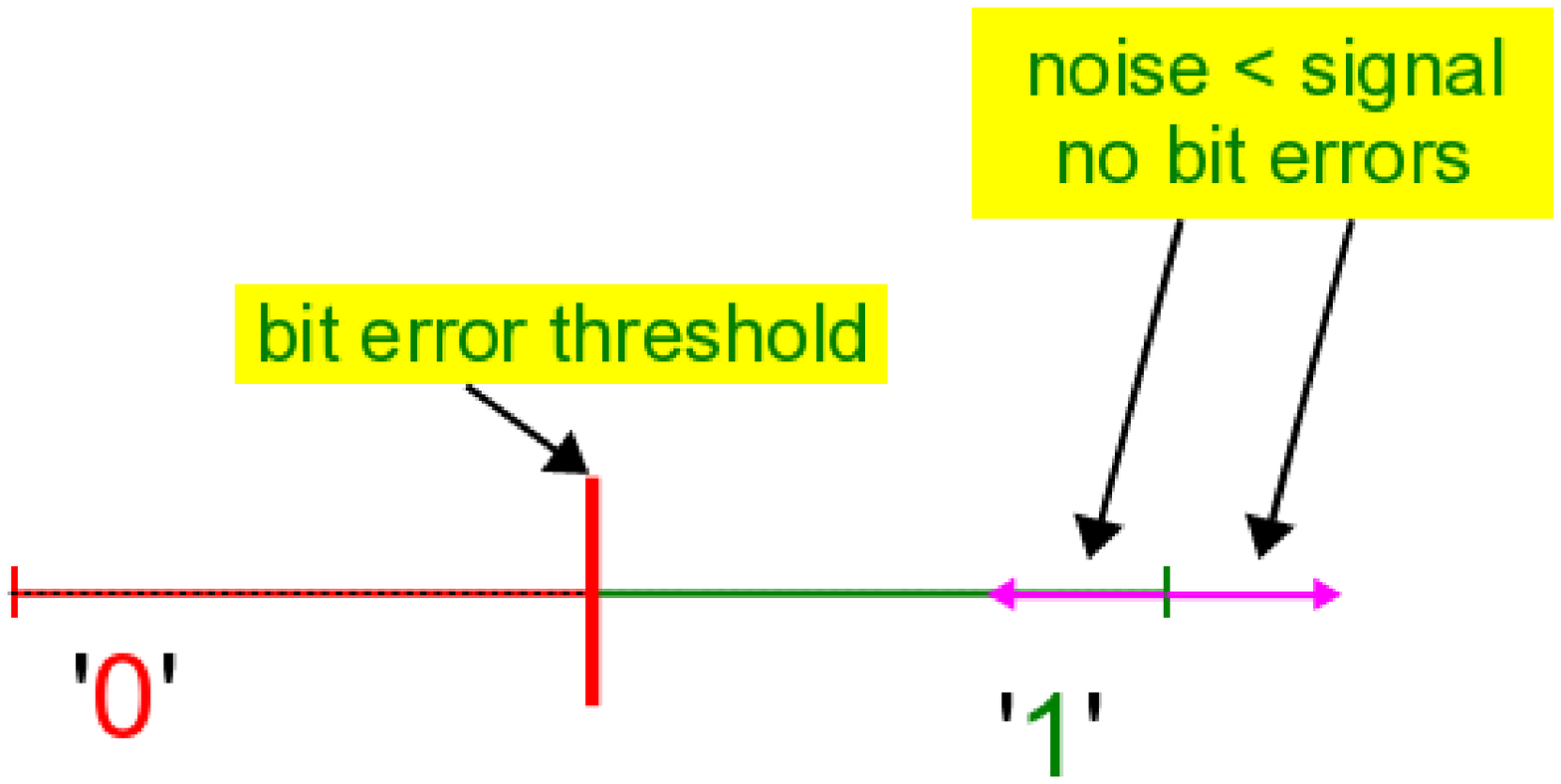,width=0.4\linewidth}
        }
        \subfigure[Large noise amplitude.\figlabel{levelbasednoiseLarge}]{
            \raisebox{0em}{\epsfig{file=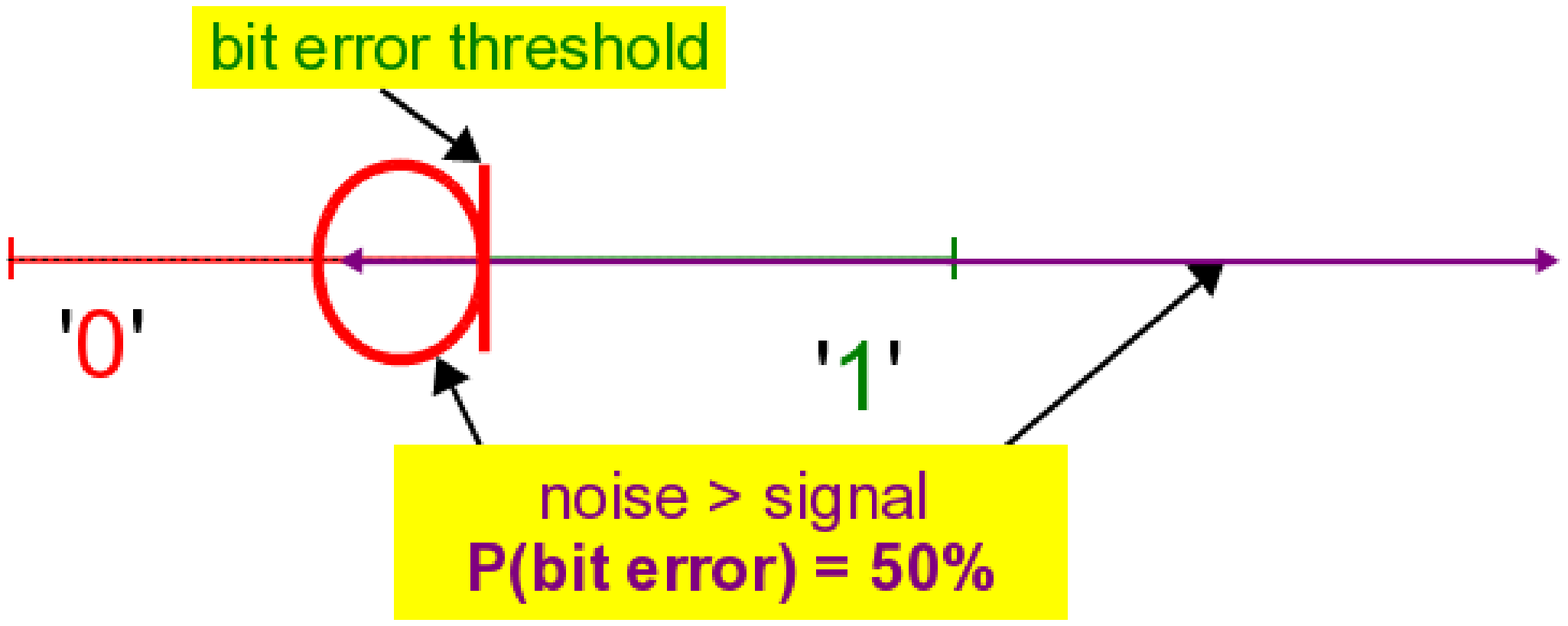,width=0.5\linewidth}}
        }
    }
    \caption{Level-based logic encoding: bit error rates for small and large noise
             amplitudes.\figlabel{levelbasednoise}}
\end{figure}

Phase-encoded logic also
offers intrinsic noise immunity advantages over level-based logic. The
underlying mechanism behind this noise immunity is easy to appreciate
graphically.

\figref{levelbasednoise} depicts the impact of small and large noise if logic is
encoded as levels. For comparison with the phase-encoded case below, a diagram
similar to \figref{phaseencodingPhasor} is used to represent the logical states
0 and 1, but these simply represent levels (with no phase); a positive level
represents 1 and a negative level (of equal amplitude) represents 0. The bit
error threshold in the presence of noise is zero.  In
\figref{levelbasednoiseSmall}, the impact of adding fixed-amplitude ``small''
noise (\ie, the noise is less than the signal) is shown. This random noise adds
to, or subtracts from, the signal with equal probability. In either case, the
resulting signal remains positive since the noise is small, hence there
is no bit error. But if the fixed noise is larger in value than the
signal amplitude, as shown in \figref{levelbasednoiseLarge}, this situation
changes. When the noise adds to the signal, there is no bit error; but when it
subtracts, there is \textit{always} a bit error, since the result becomes
negative, crossing the bit error threshold. Hence, when the noise is
larger than the signal, level-based logic encoding suffers a 50\% probability of
error, \ie, the bit becomes perfectly random, losing all information.

\begin{figure}[htbp]
    \centering{
        \subfigure[Small noise amplitude.\figlabel{phasebasednoiseSmall}]{
            \epsfig{file=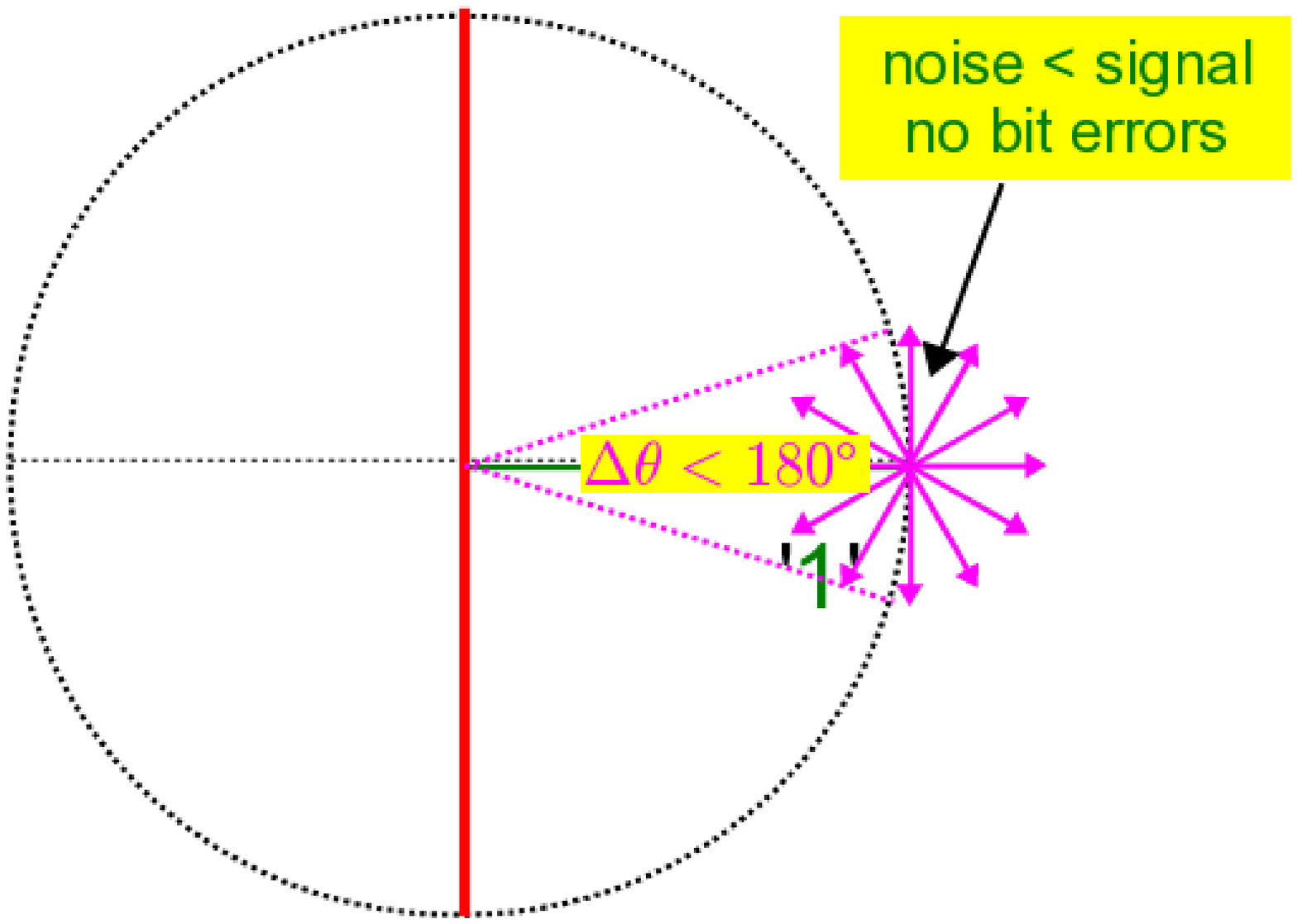,width=0.4\linewidth}
        }
        \subfigure[Large noise amplitude.\figlabel{phasebasednoiseLarge}]{
            \raisebox{0em}{\epsfig{file=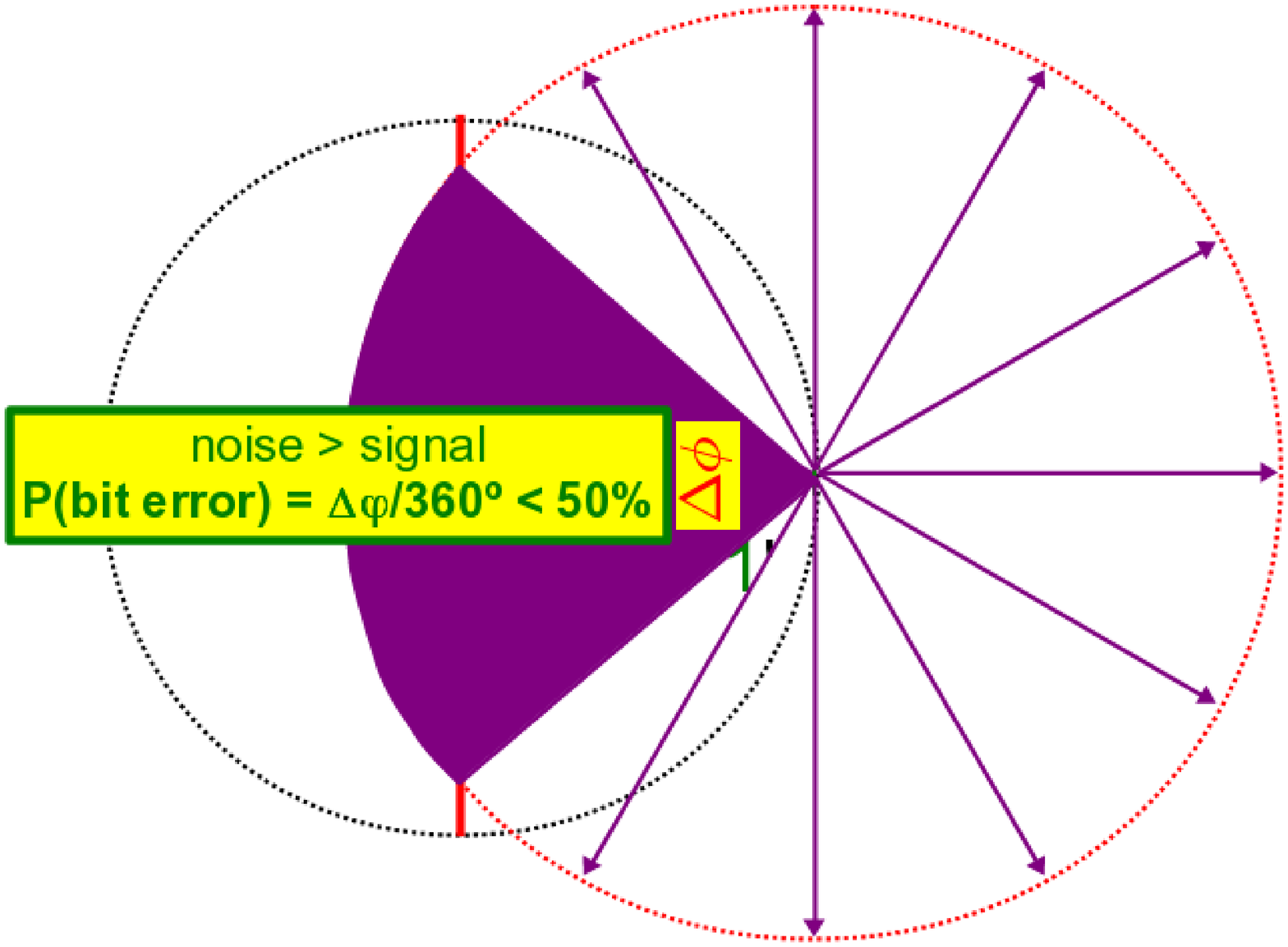,width=0.5\linewidth}}
        }
    }
    \caption{Phase-based logic encoding: bit error probabilities for small and large noise
             amplitudes.\figlabel{phasebasednoise}}
\end{figure}

The situation when logic is encoded in phase is depicted in
\figref{phasebasednoise}. Here, the signal values `0' and `1' are phasors,
exactly as in \figref{phaseencodingPhasor}; the noise added is also a phasors,
at the \textit{same frequency}. In
this case, the bit error thresholds are the vertical phasors at $\pm$90$^\circ$,
\ie, the phase halfway between the `0' and `1' states.
\figref{phasebasednoiseSmall} shows the case when the noise amplitude is less
than the signal's. Because the noise is random, its phase is uniformly
distributed in [0$^\circ$, 360$^\circ$], as shown. The worst-case phase error
caused by the additive noise, denoted $\Delta \theta$, is less than 90$^\circ$
in absolute value; hence there is no bit error. For ``small'' noise, therefore,
phase encoding and level encoding are identical from a bit error perspective.

When the noise amplitude is ``large'' (\ie, greater than the signal's), the situation
in the case of phase encoding differs markedly from that for level
encoding, as shown in \figref{phasebasednoiseLarge}. The shaded region depicts
the range of noise phases ($\Delta \phi$) that lead to a bit error. Importantly,
$\Delta \phi$ is always less than 180$^\circ$, implying a \textit{bit error
probability of less than 50\% even when the noise amplitude is greater than that of
the signal}. Indeed, for noise amplitudes that are only slightly greater than
the signal's, the bit error probability is very small, in stark contrast with
the level based case. Phase based encoding approaches a 50\% bit error
probability only as the noise amplitude tends to infinity. 

These noise characteristics of phase encoding are well known in communication
theory \cite{middleton}; in particular, the above reasoning is essentially
identical to that establishing the superior noise performance of BPSK (binary
phase shift keying) over BASK (binary amplitude shift keying). Phase based
logic encoding simply leverages this fact to improve noise immunity at the
physical implementation level of Boolean computing.

\section{Conclusion}
Recent developments in phase-encoded logic have made
it relevant as an alternative computational scheme for today's nanoscale
integration era. The fact that almost any SSNO can serve as a phase logic latch
implies that many new substrates for phase-based logic (such as spin-transfer
nano-oscillators (STNOs) \cite{ZeEtAl2013ultralowpowerSTNOs}) can potentially be
exploited. That energy-efficient oscillators serving as phase logic latches are
capable of switching very quickly in an energy-neutral manner, and that
phase encoding brings inherent noise immunity benefits, provide incentives for
exploring its use.

\subsection*{Acknowledgments}
The author thanks Tom Theis for discussions that motivated 
the energy and speed explorations in \secref{energy}. Tianshi
Wang designed and demonstrated ring oscillator based phase logic circuits and
design tools
\cite{WaRoUCNC2014PHLOGON,WaRoDAC2015MAPPforPHLOGON,WaAaWuYaRoCICC2015MAPP},
and provided valuable insights into Q factors of nonlinear oscillators.
Support from the National Science Foundation of the United States is gratefully
acknowledged. NSF Grant CCF-1111733
(PHLOGON) funded this work
directly, while support from NSF Grant EEC-1227020
(NEEDS) was instrumental in
enabling the Berkeley MAPP infrastructure
\cite{WaAaWuYaRoCICC2015MAPP,MAPPwebsite}, within which design tools for
SSNO-based phase logic are being developed.

\let\em=\it 

\bibliographystyle{unsrt}
\bibliography{stringdefs,jr,tianshi,PHLOGON-jr,von-Neumann-jr}

\end{document}